\documentclass[11pt]{amsart}
\usepackage{setspace}
\usepackage[english]{babel}
\usepackage{amsmath, amsthm}
\usepackage{amsfonts}
\usepackage{amsaddr}
\usepackage{mathtools}
\usepackage{amssymb}
\usepackage{verbatim} 
\usepackage{float}
\usepackage[pdftex]{graphicx}
\usepackage{epstopdf}
\usepackage{float}
\usepackage{caption}
\usepackage{subcaption}
\usepackage{bm}
\usepackage[hyperref]{xcolor}
\usepackage{tikz}
\usepackage[margin=1.3in]{geometry}
\usepackage{pgfplots}
\usetikzlibrary{shapes.geometric}
\usetikzlibrary{plotmarks}
\usepackage{url}
\usepackage[shortlabels]{enumitem}
\usepackage{hyperref}
\usepackage{cleveref}
\crefname{figure}{Figure}{Figures}
\crefname{exmp}{example}{examples}
\crefname{equation}{equation}{equations}
\crefname{appendix}{appendix}{appendices}

\usepackage{csquotes} 

\usepackage[
	sortcites,
	style=nature,
	intitle=true,
	giveninits=true,%
	backend=biber,
	doi=false,
	isbn=false,
	url=true,
	citestyle=numeric,
	urldate=long]{biblatex}	
\AtEveryBibitem{%
  \clearfield{month}
  \ifentrytype{book}{
  \clearfield{pages}%
  }{
  }%
  \ifentrytype{misc}{
  }{
    \clearlist{language}
    \clearfield{url}%
    \clearfield{urldate}%
  }%
}

\usepackage{algorithm}
\usepackage{enumerate}
\usepackage[noend]{algpseudocode}
\newcommand*\Let[2]{\State #1 $\gets$ #2}
\algrenewcommand\algorithmicrequire{\textbf{Input:}}
\let\oldReturn\Return
\renewcommand{\Return}{\State\oldReturn}

\newcommand{\N}{\ensuremath{\mathbb{N}}}

\newcommand{\vect}[1]{\boldsymbol{\mathbf{#1}}}

\newcommand{\id}{\,\mathrm{d}}

\newcommand{\Ex}[1]{\ensuremath{\mathbb{E}\left[ #1 \right]}}

\newtheorem*{lemma*}{Lemma}

\theoremstyle{definition}

\definecolor{blue}{rgb}{0.2980392156862745, 0.4470588235294118, 0.6901960784313725}
\definecolor{green}{rgb}{0.3333333333333333, 0.6588235294117647, 0.40784313725490196}
\definecolor{red}{rgb}{0.7686274509803922, 0.3058823529411765, 0.3215686274509804}
\definecolor{purple}{rgb}{0.5058823529411764, 0.4470588235294118, 0.6980392156862745}
\definecolor{yellow}{rgb}{0.8, 0.7254901960784313, 0.4549019607843137}
\definecolor{cyan}{rgb}{0.39215686274509803, 0.7098039215686275, 0.803921568627451}
\definecolor{bb1}{rgb}{0.0,  0.44705882,  0.69803922}
\definecolor{bb2}{rgb}{0.0,  0.61960784,  0.45098039}
\definecolor{bb3}{rgb}{0.83529412,  0.36862745,  0.0}
\definecolor{bb4}{rgb}{0.8,  0.4745098 ,  0.65490196}
\definecolor{bb5}{rgb}{0.94117647,  0.89411765,  0.25882353}
\definecolor{bb6}{rgb}{0.3372549 ,  0.70588235,  0.91372549}
\definecolor{orange}{rgb}{0.83529412,  0.36862745,  0.0}
\definecolor{oxfordblue}{rgb}{0.00000,  0.12941,  0.27843}

\graphicspath{{Graphics/}} 
\newlength\figureheight 
\newlength\figurewidth 
\hypersetup{
    linkcolor = oxfordblue,
    anchorcolor = oxfordblue,
    citecolor = oxfordblue,
    filecolor = oxfordblue,
    urlcolor = oxfordblue,
    colorlinks  = true,
}
\newif\ifhighlight
\highlightfalse

\newcommand{\highlight}[1]{\ifhighlight \textbf{#1}\else #1\fi}

\usepackage{lineno}

\title[QMC methods for tau-leaping]{Quasi-Monte Carlo methods applied to tau-leaping in stochastic biological systems}
\author[ C.~H.~L.~Beentjes and R.~E.~Baker]{Casper H.~L.~Beentjes  and Ruth~E.~Baker}
\address{Mathematical Institute, University of Oxford, Oxford, UK}
\email{\href{mailto:beentjes@maths.ox.ac.uk}{beentjes@maths.ox.ac.uk}}

\begin{document}
\begin{abstract}
Quasi-Monte Carlo methods have proven to be effective extensions of traditional Monte Carlo methods in, amongst others, problems of quadrature and the sample path simulation of stochastic differential equations. By replacing the random number input stream in a simulation procedure by a low-discrepancy number input stream, variance reductions of several orders have been observed in financial applications. 

Analysis of stochastic effects in well-mixed chemical reaction networks often relies on sample path simulation using Monte Carlo methods, even though these methods suffer from typical slow $\mathcal{O}(N^{-1/2})$ convergence rates as a function of the number of sample paths $N$. This paper investigates the combination of (randomised) quasi-Monte Carlo methods with an efficient sample path simulation procedure, namely $\tau$-leaping. We show that this combination is often more effective than traditional Monte Carlo simulation in terms of the decay of statistical errors. The observed convergence rate behaviour is, however, non-trivial due to the discrete nature of the models of chemical reactions. We explain how this affects the performance of quasi-Monte Carlo methods by looking at a test problem in standard quadrature.
\end{abstract}

\maketitle

\section{Introduction}
In the last few decades, research in molecular biology has generated vast amounts of quantitative data. This growing amount of data has inspired the development of a variety of mathematical modelling and simulation techniques aiming to support the experimental study of the intricate processes taking place in cells and other molecular systems. As a result, we now often have the option to perform \textit{in silico} experiments alongside the more traditional \textit{in vivo} and \textit{in vitro} approaches to study complex cellular pathways, allowing us to perform model fitting and inference, thus yielding a detailed view of the different components in these often intricate networks \cite{Wilkinson2009}. 

One feature which nowadays appears prominently in many models of chemical reaction networks is randomness. The aim of including randomness is to mimic the effects of intrinsic and extrinsic noise sources present in molecular systems, as found in experiments \cite{McAdams1999,Elowitz2002,Blake2003,Cai2006}. Stochasticity can be responsible for a wide variety of observed phenomena, such as stochastic focusing \cite{Paulsson2000} or resonance-inducing oscillations \cite{Hou2003}. The addition of noise, however, comes at a price in terms of our \textit{in silico} experiments. Single experiments have to be run many times using Monte Carlo (MC) methods to yield results in the form of summary statistics to a satisfactory degree of certainty. This requirement can result in large computation times or even make a problem intractable with existing computational methods. 

As such, a key component in the development of computational techniques for these stochastic models is finding ways to reduce the variance of the statistics returned by the simulation algorithms used. In other fields that rely heavily on MC computations, such as computational finance, a commonly applied variance reduction approach involves the use of quasi-Monte Carlo (QMC) techniques. By changing the random number input from pseudo-random numbers to low-discrepancy numbers a gain in performance of sometimes several orders can be achieved. In the context of the simulation of chemical reaction networks this idea has received very little attention. In \cite{Hellander2008} an exploration of the combination of QMC methods with the exact simulation of continuous time Markov chain (CTMC) models is presented which shows some of the benefits of QMC over standard methods. In this work we specifically look at an approximate simulation method, $\tau$-leaping, which allows for an easier incorporation of QMC methods. We show that the benefits from using QMC methods in this case are perhaps less striking than anticipated based on the success of QMC methods in the numerical solution of stochastic differential equations (SDEs). We then give a detailed explanation for why this is the case, which serves as an explanation for the observations made in \cite{Hellander2008} as well.

\subsection*{Outline}
We start in \Cref{sec:chemreac} with an overview of mathematical modelling of chemical reactions of well-mixed species. These models need different simulation methods to compute summary statistics, which are discussed next, in \Cref{sec:simulation}. This section also includes a discussion about MC methods and the resulting statistical error in summary statistics. To reduce the statistical error we explore the use of QMC methods, and we provide a practitioners introduction to QMC in \Cref{sec:QMC}. In \Cref{sec:results} we show results from the application of QMC methods in the simulation of chemical reaction networks. We compare the results with a simple toy model from classical quadrature to elucidate the difference in performances observed.  Finally, we conclude, in \Cref{sec:discussion} with a discussion of our observations on the combination of QMC methods and the simulation of stochastic biological systems.

\section{Mathematical models of chemical reactions}\label{sec:chemreac}
For the main part of this work, we will look at models that describe the temporal evolution of molecule copy numbers where we assume that the molecules are spatially distributed in a homogeneous way, i.e.\ we assume the systems to be well-mixed in a volume $V$. Note that it is possible to include the spatial movement of molecules in this framework and we refer to \cite{Erban2007} for more detail.

Suppose we have a collection of $n$ types of chemical species $S_1,\dots,S_n$ that can interact via $K$ different types of reactions $R_1,\dots,R_K$, often denoted as reaction channels. In generic form we can describe such an interaction $R_k$ as
\begin{equation*}
\alpha_{1,k} S_1 + \dots + \alpha_{n,k} S_n \xrightarrow{c_k} \beta_{1,k} S_1 + \dots + \beta_{n,k} S_n,\qquad k=1,\dots,K,
\end{equation*}
where $\alpha_{i,k},\beta_{i,k}\in \N$ and $c_k$ is the reaction rate constant for reaction channel $R_k$. We define $\vect{X}(t)$ to be the state vector, describing the evolution of all the species as time evolves, i.e.\ $\vect{X}_i(t)$ is the number of molecules $S_i$ at time $t$. Upon the firing of reaction $R_k$ the number of molecules can change and we will use $\vect{\zeta} _k$, the stoichiometric vector for reaction $R_k$, to denote the change in the copy number when reaction $R_k$ fires, i.e.\ due to reaction $R_k$ we see $\vect{X}\to \vect{X} + \vect{\zeta}_k$. Or, in terms of the $\alpha_{i,k}$ and $\beta_{i,k}$, we thus have the $i$-th component of $\vect{\zeta}_k$ equal to $\beta_{i,k}-\alpha_{i,k}$.  We can now describe the temporal evolution of the chemical species using
\begin{equation}
\vect{X}(t) = \vect{X}(0) + \sum_{k=1}^K N_k(t) \vect{\zeta}_k,
\end{equation}
where $N_k(t)$ denotes the number of times that reaction channel $R_k$ fires in the time interval $[0,t)$. This is, of course, not insightful yet as we have not described a means to calculate the $N_k(t)$. In order to model these counting functions, $N_k$, we will assign to every reaction channel, $R_k$, a propensity function, $a_k(\vect{X}(t))$, which describes the probability that the reaction channel fires in the infinitesimal time interval $[t,t+\id t)$. Note that the selection of this function is a modelling choice. A commonly employed choice is the \textit{Law of Mass Action} which, in essence, looks at the number of combinations of reactants that can be made using the state vector $\vect{X}(t)$ to let reaction $R_k$ fire, see for example \cite{Higham2008} (but note that the choice of $a_k$ is in no way essential to what follows).

We will now present an inverse historical way to define $N_k(t)$ based on these propensity functions. An accurate way to model this counting function would be to view the above description as a CTMC where, given the current state, $\vect{X}(t)$, we can experience $K$ different state transitions based on the various reaction channels. An inhomogeneous Poisson process, $Y_k$, will then describe the number of times reaction $R_k$ fires. This leads to the Kurtz \textit{random time change representation} (RTCR) \cite{Anderson2011}
\begin{equation}\label{eq:KurtzRT}
\vect{X}(t) = \vect{X}(0) + \sum_{k=1}^K Y_k\left(\int_0^t a_k(\vect{X}(s)) \id s\right) \vect{\zeta}_k,
\end{equation}
where we see $K$ independent inhomogeneous Poisson counting processes $Y_k$. This representation of the CTMC is a pathwise description of the dynamics. Alternatively, one can describe the CTMC by the \textit{chemical master equation} (CME), a (high) dimensional system of ordinary differential equations (ODEs) that describes the time evolution of the probability to occupy parts of the state space of $\vect{X}(t)$. Note that the CME can only be solved in special cases \cite{Jahnke2006}. Both the RTCR and the CME form the basis for many simulation approaches, some of which we will touch upon later.

Going back to \cref{eq:KurtzRT} note that we can rewrite  the time evolution of the CTMC as
\begin{equation}
\vect{X}(t+\tau) = \vect{X}(t) + \sum_{k=1}^K Y_k\left(\int_t^{t+\tau}  a_k(\vect{X}(s)) \id s\right) \vect{\zeta}_k,
\end{equation}
which now describes the evolution of the system over a time interval $[t,t+\tau)$. If we assume that $\tau$ is small enough such that $a_k(\vect{X}(s))\approx a_k(\vect{X}(t))$ 
over the interval $[t,t+\tau)$, but still such that $a_k(\vect{X}(t))\tau\gg 1$, we can use two approximations to derive the \textit{chemical Langevin equation} (CLE) \cite{Gillespie2000}. First we note that because $a_k(\vect{X}(s))\approx a_k(\vect{X}(t))$ we can write
\begin{equation}\label{eq:pre-tau}
\vect{X}(t+\tau) \approx \vect{X}(t) + \sum_{k=1}^K Y_k\left(a_k(\vect{X}(t)) \tau\right) \vect{\zeta}_k,
\end{equation}
where we now have $K$ homogeneous Poisson processes. Next we use the normal approximation of a Poisson process with large rate parameter $a_k(\vect{X}(t))\tau\gg 1$ to give
\begin{align}\label{eq:Langevin}
\vect{X}(t+\tau) &\approx \vect{X}(t) + \sum_{k=1}^K \mathcal{N}_k\left( a_k(\vect{X}(t)) \tau, a_k(\vect{X}(t)) \tau \right)\vect{\zeta}_k \nonumber \\
&= \vect{X}(t) + \sum_{k=1}^K \left[ a_k(\vect{X}(t))\tau + \sqrt{a_k(\vect{X}(t)) \tau} \mathcal{N}_k\left(0,1\right) \right] \vect{\zeta}_k,
\end{align}
where $\mathcal{N}_k(\mu,\sigma^2)$ is a normal random variable with mean $\mu$ and variance $\sigma^2$. \Cref{eq:Langevin}, in the limit $\tau\to 0$, gives an evolution equation for $\vect{X}(t)$ in the form of an SDE, which can be written in the form
\begin{equation}\label{eq:cle}
\id \vect{X}_t = \left[ \sum_{k=1}^K a_k(\vect{X}_t) \vect{\zeta}_k\right]\id t +  \sum_{k=1}^K \sqrt{a_k(\vect{X}_t)}\vect{\zeta}_k\id W_{t,k},
\end{equation}
where now the $W_{t,k}$ denote $K$ independent Wiener processes. Comparing the RTCR, \cref{eq:KurtzRT}, with \cref{eq:cle} we see that both are pathwise descriptions of the dynamics of the species. In the case of the CLE it is also possible to describe the dynamics in a manner akin to the CME, i.e.\ in terms of occupation probability of the state space of $\vect{X}(t)$. This approach yields the \textit{chemical Fokker-Planck equation} (CFPE), a system of partial differential equations (PDEs) that, in general, again is high-dimensional, and not many systems that yield exact results are known.

Using the SDE formulation we can finally derive the deterministic \textit{reaction rate equations} (RREs) which have been in use for over a century. We take the thermodynamic limit, letting the number of molecules and the volume $V$ go to infinity whilst their ratio remains constant. In this limit the random fluctuations become negligibly small compared to the deterministic terms and we convert \cref{eq:Langevin} into
\begin{equation}
\vect{X}(t+\tau) = \vect{X}(t) + \sum_{k=1}^K a_k(\vect{X}(t)) \tau \vect{\zeta}_k,
\end{equation}
which we can rewrite into a system of ODEs by taking the limit $\tau\to 0$:
\begin{equation}\label{eq:rateODE}
\frac{\id \vect{X}(t)}{\id t}= \sum_{k=1}^K a_k(\vect{X}(t)) \vect{\zeta}_k.
\end{equation}
Because they constitute a relatively smaller system of ODEs, the RREs can be studied using analytical tools, or numerical ODE solvers. On the flip side, however, this system of ODEs is completely deterministic and therefore \highlight{does not incorporate stochastic effects}. 

We thus have three different mathematical models describing the temporal evolution of well-mixed molecular species $S_1,\dots,S_N$, namely the RTCR \eqref{eq:KurtzRT}, the CLE \eqref{eq:Langevin} and the RREs \eqref{eq:rateODE}, respectively. These models can be seen as a chain of approximations going from a CTMC with discrete state space to a CTMC with a continuous state space to a deterministic ODE system. Each of these models can be analysed and simulated in different ways and at different cost. 

\section{Simulation of well-mixed systems}\label{sec:simulation}
Having presented three different models for the evolution of interacting chemical species in the preceding section we can now ask how to perform a mathematical analysis on them. For the deterministic rate equations \eqref{eq:rateODE} we can use a wide array of well-known analytical and numerical techniques: we will not go in detail here but rather refer to \cite{Higham2008} and references therein. If, however, noise and non-linear reactions (for example second order or higher with mass action) are important the RRE model will not yield correct results and one has to resort to one of the two stochastic models mentioned previously to investigate system behaviour.

The formulations using CTMCs incorporate stochasticity and therefore are often harder to interrogate using analytic methods; only for a handful of cases this has proven to be possible \cite{Jahnke2006}. Repeated simulation of sample paths from these models is therefore crucial in order to gain insight into the dynamics of the species numbers. For the CLE we can use standard simulation procedures that are used for SDEs, such as the Euler-Maruyama (EM) or Milstein methods, and we refer to \cite{Kloeden1992} for an extensive exposition of that subject. These computational techniques for SDEs are generally efficient compared to methods for the RTCR that we will discuss next. However, they do suffer from both a numerical error depending on the integration scheme used and a bias error, because the CLE \eqref{eq:cle} is only an approximation to the RTCR \eqref{eq:KurtzRT}.

The dynamics of the RTCR \eqref{eq:KurtzRT} can generally be studied in two different ways. The first is by use of the CME. However, one of the disadvantages of this approach is the dimension of this system of ODEs; it is equal to the size of the state space. This will generally be so high that the problem becomes intractable, and we refer to \cite{Schnoerr2017} for a general overview of computational approximations related to the CME.

This means that one is often forced to rely on a different approach to get a handle on the model dynamics. Instead of looking at the evolution of the probability over the whole state space at once we generate single sample paths which evolve according to the rules of \cref{eq:KurtzRT}. An exact algorithm to compute such sample paths in the context of chemical reaction networks is called the Stochastic Simulation Algorithm (SSA) or Gillespie's Direct Method \cite{Gillespie1977}. Given a current state $\vect{X}(t)$ the algorithm 
provides a way to compute the time until the next reaction fires and determines which reaction this is. In that way we can progress the Markov chain one reaction at a time. This approach can be made more computationally efficient by, for example, using the Next Reaction Method by Gibson and Bruck \cite{Gibson2000} or the Modified Next Reaction Method by Anderson \cite{Anderson2007}. 

%


Still these methods suffer from a drawback, namely their computational costs. As they simulate each reaction individually their run time can be significant if we have many molecules and reactions involved. This is the rationale behind the development of approximate methods to simulate sample paths from \cref{eq:KurtzRT}. One of the most widely used methods is the $\tau$-leap scheme, also developed by Gillespie \cite{Gillespie2001}. We go back to \cref{eq:pre-tau}, but this time we do not approximate the Poisson process by Gaussian random variates to yield the CLE. In essence the $\tau$-leap method follows from the rationale that, given a small enough $\tau$, the propensities of the reactions do not change much in the time interval $[t,t+\tau)$ and therefore can be assumed constant. This approach yields a discrete-time Markov chain (DTMC) with a discrete state space, where the time between each state is given by the time-step $\tau$ and the transitions are computed by
\begin{equation}\label{eq:tau-leap}
\vect{X}(t+\tau) = \vect{X}(t) + \sum_{k=1}^K Y_k\left(a_k(\vect{X}(t)) \tau\right) \vect{\zeta}_k.
\end{equation}
The computational gain with this method is that in order to calculate $Y_k\left(a_k(\vect{X}(t)) \tau\right)$ we can simply generate a single Poisson random variable $p_k$ with parameter $a_k(\vect{X}(t)) \tau$. This means that we can fire multiple reactions at once and therefore progress quicker than is the case for the SSA. An algorithmic representation of the $\tau$-leap method is depicted in \Cref{algo:tau-leap}.

\begin{algorithm}
\caption{$\tau$-leap method}
\label{algo:tau-leap}
\begin{algorithmic}[1]
\Require{Initial data $\vect{X}(0)=\tilde{\vect{X}}$}
\Require{Stoichiometric matrix $\vect{\zeta}$}
\Require{Propensity functions $a_k(\vect{x})$}
\Require{Time step $\tau$}
\Require{Final time $T$}

\Let{$\vect{X}$}{$\tilde{\vect{X}}$}
\Let{$t$}{0}
\While{$t<T$}
\Let{$A_k$}{$a_k(\vect{X})$} \Comment{Calculate the propensities.}
\State{Generate $p_1,\dots,p_K$ Poisson random variables with intensity parameters $A_1,\dots,A_k$.}
\Let{$\vect{X}$}{$\vect{X}+\sum_{k=1}^K \vect{\zeta}_k p_k.$} \Comment{Update state vector.}
\Let{$t$}{$t+\tau$} \Comment{Update time.}
\EndWhile
\Return{$\vect{X}$}
\end{algorithmic}
\end{algorithm}

Being an approximate method the $\tau$-leap method does not come without caveats, one being the possibility of achieving negative molecule numbers. Many possible workarounds to avoid negative copy numbers have been proposed \cite{Tian2004,Chatterjee2005,Cao2005,Cao2006}. Furthermore, because $\tau$-leaping is an approximate method, it yields a bias depending on selection of the magnitude of the step size, $\tau$ \cite{Anderson2011a}. Therefore one has to balance the effect of this bias with the computational costs, which are $\mathcal{O}(\tau^{-1})$.

A different view on the $\tau$-leap method is that it is a variant of the explicit Euler method for ODEs applied to \cref{eq:KurtzRT}, where we have approximated the time integral by a left Riemann sum. This method therefore parallels the widely used EM scheme for SDEs and one could therefore ask the question whether it is possible to adapt other ODE time stepping approaches to the CTMC simulation case. Indeed it is possible for a small class of methods such as implicit Euler, which yields implicit $\tau$-leap approaches \cite{Rathinam2003} and these methods can perform better for systems exhibiting e.g.\ stiff behaviour. 

\subsection*{Monte Carlo methods and errors}
As a final note on the simulation of well-mixed systems we now reflect on the different methods mentioned above. Many commonly used methods provide means to generate (approximate) sample paths of chemical reaction networks, but how can we infer information from these? In many instances we are interested in expressions like $g(\vect{X}(t))$, where $g$ is a function of the state variable $\vect{X}(t)$ at time $t$. However, as $\vect{X}(t)$ is a random variable we will often have to look at the expectation $\Ex{g(\vect{X}(t))}$ of this function $g$. A common example would be $g(x)=x^k$, where taking the expectation yields the $k$-th moment of the process $\vect{X}(t)$.

If we can only generate sample paths from the distribution of possible outcomes in the state space we have to employ MC methods to estimate the required statistics. We generate $N$ independent, possibly approximate, sample paths $\hat{\vect{X}}^{(1)}(t),\dots,\hat{\vect{X}}^{(N)}(t)$ and construct from this the MC estimator
\begin{equation}
E_N(t)=\frac{1}{N}\sum_{n=1}^N g\left(\hat{\vect{X}}^{(n)}(t)\right)\approx \Ex{g\left(\vect{X}(t)\right)}.
\end{equation}
\highlight{
The more samples $N$ used, the more certain one can be of the closeness of $E_N$ to the expected value $\Ex{g(\vect{X}(t))}$. We can make this precise by considering the mean squared error (MSE) given by
}
\begin{equation}\label{eq:MSE}
\text{MSE}(E_N) = \mathbb{E}\left[\left(E_N - \Ex{g(\vect{X}(t))}\right)^2\right] =  \Big(\underbrace{\mathbb{E}\left[ E_N \right] - \mathbb{E}\left[g(\vect{X}(t))\right]}_{\text{bias}}\Big)^2 + \underbrace{\mathbb{V}\left[E_N \right]}_{\text{statistical error}},
\end{equation}
which can be decomposed into two separate sources of error. First of all there can be a bias. This could be the result of a modelling choice, for example the CLE and its related computational methods form a biased approximation of the RTCR from which it was derived. Alternatively the bias could stem from algorithm parameters such as the time step $\tau$ in the $\tau$-leap method. This type of error will not be investigated in this paper and we will assume that, for a specific computational method, it is given and fixed. 
\highlight{
On the other hand there are statistical errors, in the form of the variance $\mathbb{V}\left[E_N\right]$ of the estimator, which can be controlled. Statistical errors will therefore form the main focus of interest in this manuscript and are discussed next.} 

\highlight{It goes without saying that it is desirable to have this statistical error as small as possible; we aim to control statistical uncertainty as tightly as possible given our computational resources.} For a standard MC method, \highlight{given} a sample variance $\sigma^2$, which is determined by the model being studied, the variance of the estimator $E_N$ is given by $\sigma^2/N$. Reducing the statistical error can now be done in two ways. Firstly, by taking more samples (the variance decays to zero as $N\to\infty$). This approach requires the development of more efficient algorithms in order to bring the cost per simulation down. Alternatively, one could hope to reduce the sample path variance, $\sigma$, by applying a variance reduction technique, see \cite[Chapter 4]{Lemieux2009} for more detail in the context of MC methods. Note that employing standard variance reduction techniques results in a smaller $\sigma$ and therefore these techniques only improve the constant of convergence for $\mathbb{V}\left[E_N\right]$, the convergence rate behaviour as $N\to\infty$ does not change. For the remainder of the manuscript, however, we look at a variance reduction method different from MC that aims to reduce the variance decay as a function of increasing $N$.

\section{Quasi-Monte Carlo methods}\label{sec:QMC}
\begin{figure}[h!]
\includegraphics[width=0.6\textwidth]{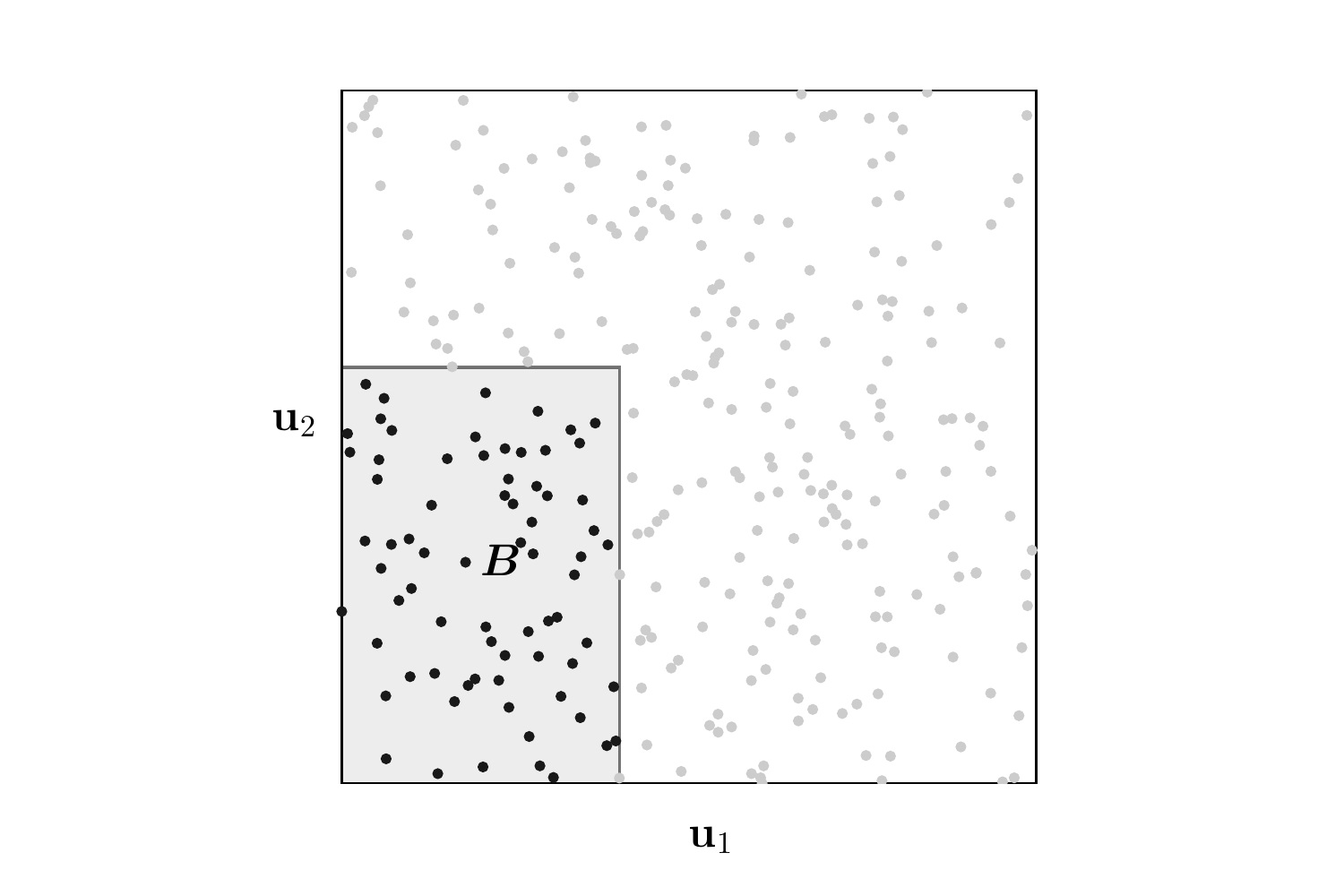}
\caption{Illustration of the discrepancy concept in $[0,1)^2$. Shown are $N=300$ points $\left\lbrace\vect{u}^{(i)}\right\rbrace$ scattered at random. If a perfect uniform distribution was attained by these points the number of points in $\boldsymbol{B}$ would be equal to $\text{Vol}(\boldsymbol{B}) \cdot N$ and this would hold true for every box $\boldsymbol{B}\subseteq [0,1)^2$.}
\label{fig:discrepancy}
\end{figure}

One of the drawbacks of general MC methods is the slow convergence rate, often of the order $\mathcal{O}(N^{-1/2})$ for the root mean squared error (RMSE). A way to improve on plain MC methods is the use of QMC methods. Originally QMC methods were developed to approximate multidimensional integrals of the form
\begin{equation}\label{eq:integral}
I=\int_{[0,1]^s} f(\vect{x})\id \vect{x},
\end{equation}
where $s$ is the dimension of the problem. In standard MC we would generate a sequence $\vect{u}^{(i)}$ with $i=1,\dots,N$ of $s$-dimensional uniform random variates and calculate
\begin{equation}\label{eq:QMC-integral}
I_N = \frac{1}{N}\sum_{i=1}^N f(\vect{u}^{(i)})\approx \int_{[0,1]^s} f(\vect{x})\id \vect{x}.
\end{equation} 
The convergence $I_N\to I$ as $N\to\infty$ for MC methods is based on the \textit{Law of Large Numbers} (LLN), but this is not necessary for convergence. For example, deterministic quadrature methods such as the midpoint-rule exist and have no relation to the LLN. It turns out that, by virtue of the Koksma-Hlawka inequality, we can link the rate of convergence of $I_N$ as $N\to\infty$ to the uniformity of the points $\left\lbrace\vect{u}^{(i)}\right\rbrace \subset [0,1)^s$ used. More precisely, the Koksma-Hlawka inequalities link the discrepancy $D^{*}_N$ of the point set $\left\lbrace\vect{u}^{(i)}\right\rbrace$ and convergence of the approximate integral. This is given in the most common form by
\begin{equation}\label{eq:Koksma-Hlawka}
\left|I_N - I \right| \leq V[f] D_N^*,
\end{equation}
where $V[f]$ is the total variation of the integrand $f$ in the sense of Hardy and Krause. This approximation error inequality can be thought of as the equivalent of \cref{eq:MSE} for MC methods. Note that \cref{eq:MSE} is an equality and holds in probability whereas \cref{eq:Koksma-Hlawka} is a deterministic worst-case inequality. Comparing the two error bounds we see that $V[f]$ takes the place of the variance $\sigma$, both quantities depending on the integrand $f$. Furthermore we see that, rather than having an error decay like $N^{-1/2}$, we now have a factor $D_N^*$ determining the behaviour as $N$ increases. The discrepancy, or the star-discrepancy $D^{*}_N$ in particular, measures the greatest deviation of a point set from a perfect uniform distribution on $[0,1)^s$, which is illustrated in \Cref{fig:discrepancy}. Taking the supremum over all the boxes $\boldsymbol{B}$ with one corner at the origin we measure the difference between the expected number of boxes in the perfect uniform case and reality. The total variation $V[f]$ of the integrand is for all practical purposes impossible to calculate and harder to estimate than the actual integral $I$. Furthermore in practical applications one can encounter functions with infinite $V[f]$, which voids the use of \cref{eq:Koksma-Hlawka}.

\begin{figure}[h!]
\begin{subfigure}[b]{0.49\textwidth}
\includegraphics[width=\textwidth]{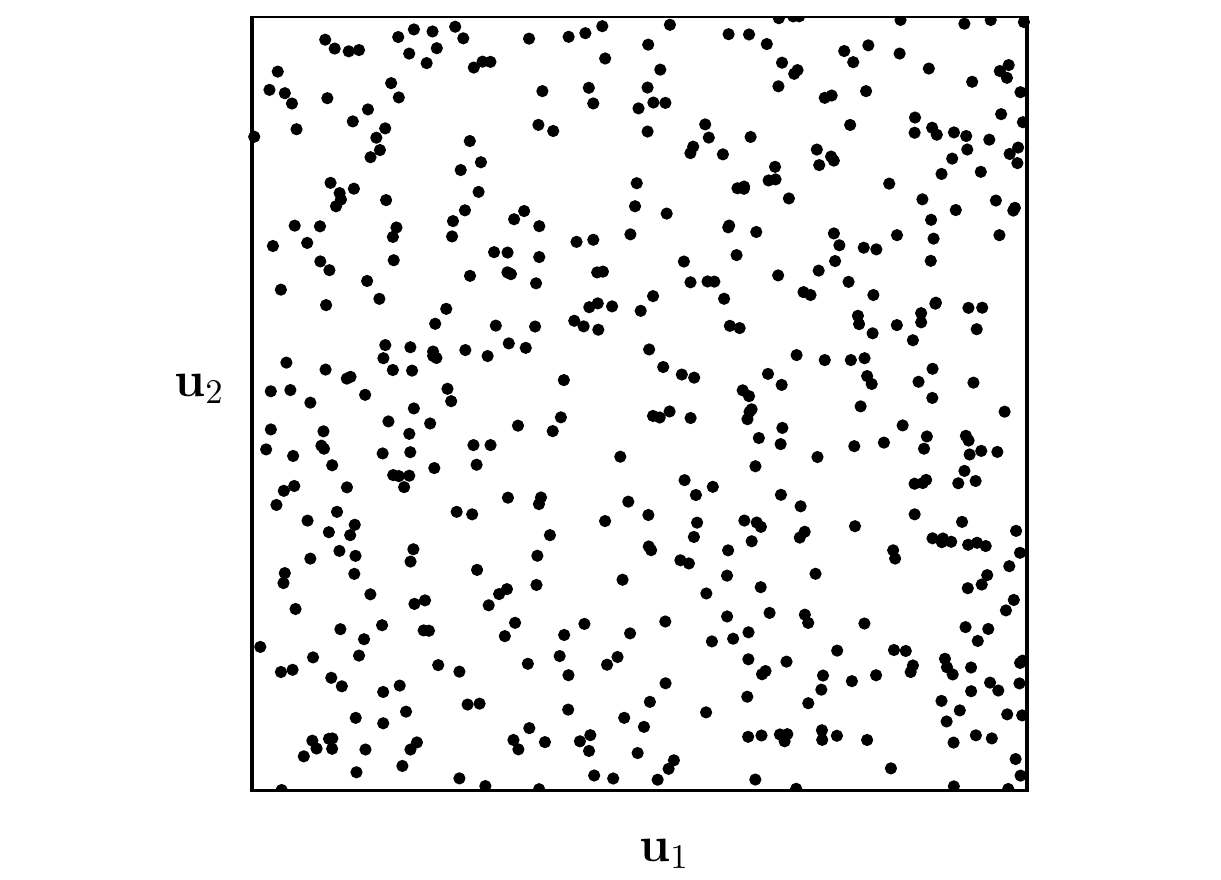}
\end{subfigure}
\begin{subfigure}[b]{0.49\textwidth}
\includegraphics[width=\textwidth]{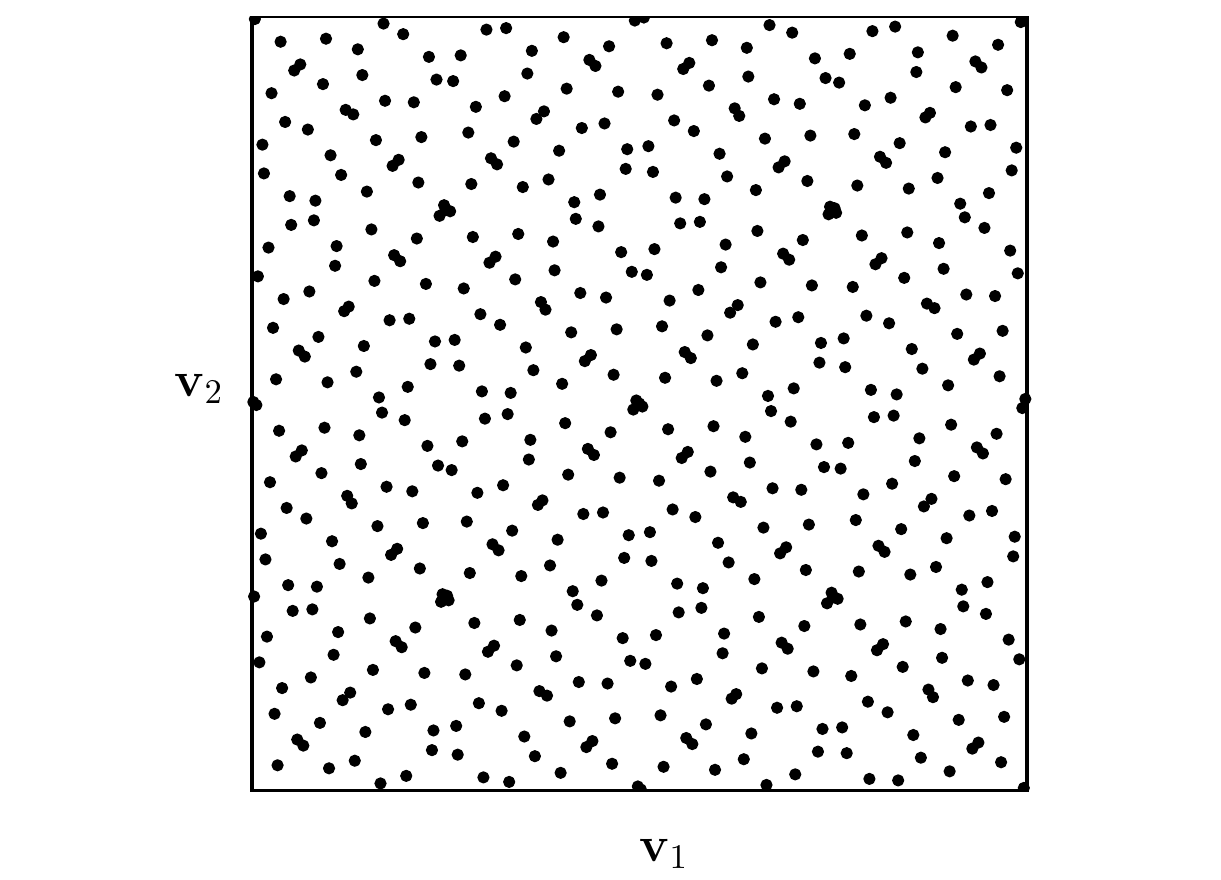}
\end{subfigure}
\caption{Comparison between a (pseudo) random point set (\textit{left}) and a low-discrepancy Sobol' point set (\textit{right}) in $[0,1)^2$, both of length $N=2^9$.}
\label{fig:pointset}
\end{figure}

It turns out that it is possible to construct low-discrepancy sequences that will cover the integration domain more uniformly than random numbers, i.e.\ their discrepancy decays quicker than for equivalent random sequences, which have $D_N^* = \mathcal{O}(N^{-1/2})$. An example comparison between a (pseudo) random sequence and a low-discrepancy sequence is depicted in \Cref{fig:pointset}, which shows that the low-discrepancy sequence attains a much better spread over the integration domain $[0,1)^2$.

For the QMC method we replace the (pseudo) random sequence $\{\vect{u}^{(i)}\}$ by a deterministic sequence of low-discrepancy numbers $\{\vect{v}^{(i)}\}$ \cite[Chapter 5]{Lemieux2009}. By their deterministic construction these sequences can attain convergence orders like $\mathcal{O}((\ln N)^s N^{-1})$ for a wide range of integrands $f$ by virtue of \eqref{eq:Koksma-Hlawka}. This $\mathcal{O}((\ln N)^s N^{-1})$ convergence rate in the limit of $N\to\infty$ will always be better than can be attained with standard MC, but if the dimension, $s$, is large but $N$ is not very large it is not clear, based on theoretical results, whether QMC will provide an improvement. There are, however, various reports in the literature, albeit without a theoretical justification, of QMC methods seemingly outperforming MC methods. \highlight{Nowadays, QMC finds its application in many more areas than just integration, such as finance, see for example \cite[Chapter 7]{Lemieux2009}, and Bayesian inference \cite{Gerber2015}. }

\subsection*{Randomised quasi-Monte Carlo}
A weakness of QMC methods compared to other quadrature methods is the lack of a measure of error. For MC methods we can use the LLN to estimate the variance and obtain confidence intervals. However, for QMC methods the points used are deterministic and therefore do not allow the application of the LLN. The Koksma-Hlawka inequality \eqref{eq:Koksma-Hlawka} does provide deterministic error bounds, but for all practical purposes the quantities involved, $V[f]$ and $D_N^*$, cannot be calculated or computed. Furthermore we note that, because the low-discrepancy numbers are a deterministic set, the QMC estimator is not unbiased. 

We can, however, consider a hybrid of MC and QMC methods. This type of approach introduces randomness into QMC methods in such a way that we keep their good convergence properties whilst at the same time allowing for error estimation with the LLN. The resulting methods are also known as randomised quasi-Monte Carlo (RQMC) methods.

A common idea in such RQMC methods is to take a low-discrepancy point set $\{\vect{v}^{(i)}\}$ and apply a randomisation to get a new set $\{\tilde{\vect{v}}^{(i)}\}$. Good randomisations (specific for the low-discrepancy sequence used) exist such that this new set is still a low-discrepancy point set but, at the same time, for all points in this set $\tilde{\vect{v}}^{(i)} \sim \mathcal{U}([0,1)^s)$ holds. As a result of such a randomisation $I_N$ with $\{\tilde{\vect{v}}^{(i)}\}$ will be an unbiased estimator of $I$. We refer to \cite{Lemieux2009} and references therein for more information on such randomisations. 

To construct a measure of the statistical error we create $M$ different randomised low-discrepancy point sets $\{\tilde{\vect{v}}^{(i,1)}\},\dots,\{\tilde{\vect{v}}^{(i,M)}\}$ which each will yield an unbiased estimator $I_{N}^{(m)}$ of the objective $I$ if we use \cref{eq:QMC-integral}. Combining these $M$ randomisations gives rise to a new estimator
\begin{equation}
I_{M,\text{RQMC}} = \frac{1}{M} \sum_{m=1}^M I_{N}^{(m)} = \frac{1}{M} \sum_{m=1}^M \left(\frac{1}{N}\sum_{i=1}^N f\left(\tilde{\vect{v}}^{(i,m)}\right)\right),
\end{equation}
which we note again is an unbiased estimator of $I$. At the same time, we can now estimate the variance like we did for MC methods, because we effectively have $M$ independent unbiased estimates of $I$. This allows for an unbiased estimator of the sample path variance just as one can obtain for standard MC simulations
\begin{equation}
\hat{\sigma}^2_{\text{RQMC}} = \frac{1}{M-1}\sum_{m=1}^M \left(I_{N}^{(m)} - I_{M,\text{RQMC}} \right)^2.
\end{equation}
We can now incorporate this into the MC framework to find an unbiased empirical estimator of $\mathbb{V}\left[I_{M,\text{RQMC}}\right]$
\begin{equation}\label{eq:var-QMC}
\sigma_{M,\text{RQMC}}^2 = \frac{\hat{\sigma}_{\text{RQMC}}^2}{M} = \frac{1}{M(M-1)}\sum_{m=1}^M \left(I_{N}^{(m)} - I_{\text{M,RQMC}} \right)^2.
\end{equation}
As a result, there are two ways one can reduce the variance of an RQMC estimator, either by taking more samples, $N$, per randomisation or by taking more randomisations, $M$. It is not always clear what choice one should make in this regard, but we can make some general observations. We note that increasing $N$ means that within each randomisation more points of the low-discrepancy set will be used. This will therefore take advantage of the better spread of low-discrepancy point sets by lowering $\hat{\sigma}_{\text{RQMC}}^2$, possibly at a rate faster than $\mathcal{O}(N^{-1/2})$. On the other hand, $M$ only controls the number of randomisations, which ties in with the standard MC framework. Therefore $M$ has limited influence on the statistical error convergence ($\mathcal{O}(M^{-1/2})$ for the RMSE). However, it should be large enough to make the variance estimation \cref{eq:var-QMC} sufficiently accurate, which can often already happen for $m\geq 10$ \cite{Lemieux2009}. Note that to get an RQMC estimator and sample variance we use $MN$ sample points and thus for a fair comparison an RQMC method should be compared to standard MC with $MN$ sample points.

In this section RQMC was introduced as a variation on standard QMC methods by adding MC style randomisations. An alternative perspective of RQMC is starting from a MC method and then adding the low-discrepancy points to make it into a variance reduction method for standard MC methods. In \Cref{ap:RQMCvarred} we give more detail on this viewpoint of RQMC.

\subsection*{Application to stochastic simulations}
(R)QMC methods were introduced in the previous sections in the context of quadrature, but the framework applies equally well to many stochastic simulation approaches. This is due to the fact that the object of interest often takes the form of an expectation, which can also be written as an integral. Therefore it can be sufficient, just as for quadrature, for stochastic simulations to substitute pseudo-random numbers in a MC simulation method with low-discrepancy numbers to get an equivalent (R)QMC method.

\begin{figure}[h!]
\includegraphics[width=0.5\textwidth]{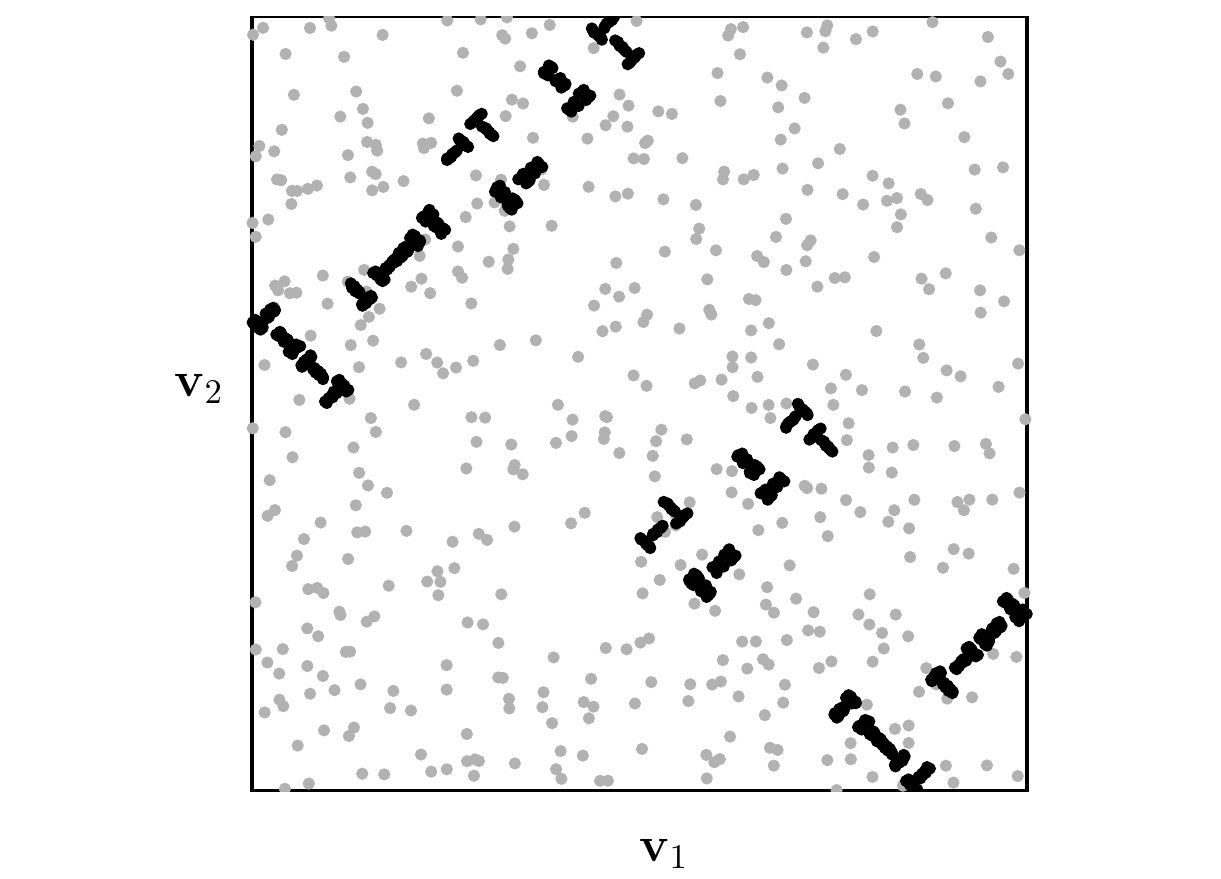}
\caption{Illustration of the combination of two one-dimensional point sets into a two-dimensional set, both for randomised Sobol' sets (\textcolor[rgb]{0,0,0}{$\bullet$}) and pseudo random sets (\textcolor[rgb]{0.7,0.7,0.7}{$\bullet$}). This approach for pseudo random numbers results in a new two-dimensional pseudo random number set, but this is not true for low-discrepancy numbers.}
\label{fig:combineset}
\end{figure}

A crucial difference, however, is that for \highlight{most standard} low-discrepancy numbers we need to know the dimensionality of the problem \textit{a priori}. This is due to the fact that one cannot make a low-discrepancy point set in two dimensions by simply combining two one-dimensional point sets (note that this does work for pseudo-random numbers!), which can be clearly seen in \Cref{fig:combineset}. This difference between the two types of points is caused by the way low-discrepancy point sets are generated, in a well-defined deterministic manner, which introduces correlation between the individual points. 

It is therefore not straightforward to combine QMC methods with e.g.\ Gillespie's SSA, as it is not clear, \textit{a priori}, how many random numbers will be used in the simulation of a single path, i.e.\ the dimension is unknown and possibly infinite.\highlight{ There do exist ways to deal with possibly infinite integration problems in the context of QMC using (extensible) lattice rules and sequences, see for example \cite[Section 5]{Dick2013} for an overview and \cite{Lecuyer2016} for a software implementation of such constructions. For chemical reactions a workaround for the simulation of CTMCs, using uniformisation of the CTMC, was presented in \cite{Hellander2008}.
}

In this paper, however, we focus instead on the (approximate) $\tau$-leaping method, which in its simplest form (fixed $\tau$) does allow for an \textit{a priori} determination of the dimension of the problem. Given $K$ reaction channels and a simulation that runs with time step $\tau$ until final time $T$ we find the dimension to be $K\lceil T/\tau\rceil$, representing the total amount of random numbers used for a single path. The low discrepancy numbers are then used in step 5 of \Cref{algo:tau-leap} to generate the Poisson random variables $p_k$ by applying an inverse transformation. Note that if this is done using a fast inverse transform, such as in \cite{Giles2016}, the process is not slower than direct methods for generating Poisson random variables in the current implementations of MATLAB and Python (R2017b and Numpy 1.14.0, respectively).

\section{Numerical experiments}\label{sec:results}
We now test the effect of the combination of RQMC and $\tau$-leaping on a set of example chemical reaction systems. We compare the results using $\tau$-leaping to the results from numerically solving the CLE \eqref{eq:cle} using the EM discretisation as QMC methods have proven to be very effective for numerical simulation of SDEs in the past \cite{Glasserman2003}. We note that the two computational methods are based on different models, the RTCR \eqref{eq:KurtzRT} and the CLE \eqref{eq:Langevin}, respectively. As a result, the bias of the methods will be different and we therefore do not directly compare the summary statistics computed. \highlight{Instead, we ignore bias and only measure the convergence rate of statistical errors for both methods. For work on the bias error incurred from using $\tau$-leaping we refer to \cite{Anderson2012a,Anderson2011a,Rathinam2016}.}

All numerical results for RQMC methods used as input the Sobol' sequences \cite{Sobol1967}, with a \highlight{random linear scramble combined with a random digital shift \cite{Matousek1998}} to create randomised low-discrepancy points.

\subsection*{Monomolecular reaction networks}
First we look at some elementary test systems to be able to closely compare the CLE-based method and the $\tau$-leap method. The benefit of these systems is that the bias due to the finite step size $\tau$ is exactly known. In addition to this the first two moments of the sample paths can be calculated analytically for both the $\tau$-leap method and the EM discretisation scheme.

\subsubsection*{Linear birth-death process}
The first example is a single species linear birth-death system
\begin{subequations}\label{eq:birth-death}
\begin{align}
S_1 &\xrightarrow{c} \emptyset,\\
S_1 &\xrightarrow{c} 2S_1,
\end{align}
\end{subequations}
which models auto-catalytic production and degradation of the species $S_1$. For simplicity we take the two reaction rates equal to each other so that we have $\Ex{\vect{X}(t)} = \vect{X}(0)$ and $\mathbb{V}\left[\vect{X}(t)\right] = 2 c t \vect{X}(0)$, i.e.\ the system will exhibit fluctuations around the steady state given by the initial state $\vect{X}(0)$. Note that these identities also hold for the EM discretisation of the CLE and the $\tau$-leap scheme applied to the RTCR, both computational methods are thus unbiased with respect to the CTMC model.

\begin{figure}[h!]
\includegraphics[width=0.65\textwidth]{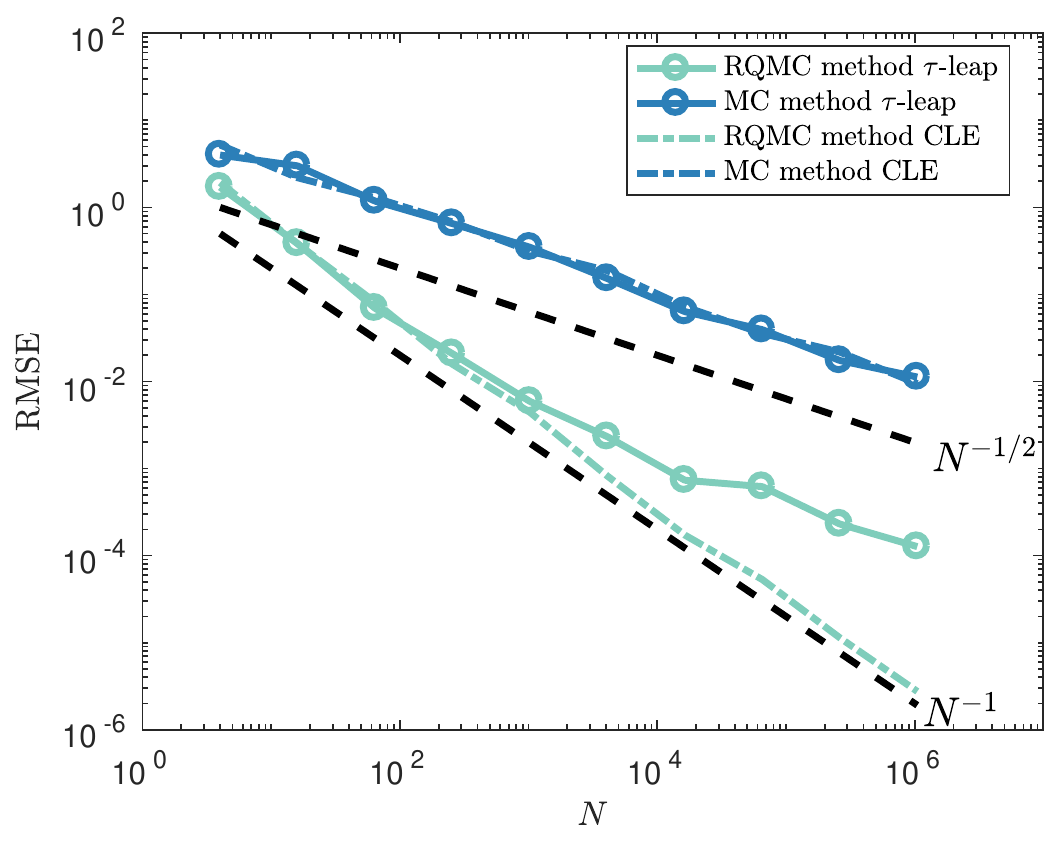}
\caption{RMSE convergence for the mean number of $S_1$ molecules in \eqref{eq:birth-death} with $c=1$ and $\vect{X}(0)=10^3$ at $T=1.6$. The time step was $\tau=0.2$ in all simulations. To establish the RMSE \cref{eq:var-QMC} was used with $M=32$ randomisations, both for the RQMC and MC methods. Dotted lines show the typical reference convergence rates, $\mathcal{O}(N^{-1/2})$  for MC and $\mathcal{O}(N^{-1})$ for RQMC.}
\label{fig:autocatalytic}
\end{figure}

In \cref{fig:autocatalytic} we show the convergence results of the RMSE at time $T=1.6$ for a system with $c=1$ and $\vect{X}(0)=10^3$. Both the Euler-Maruyama discretisation of the CLE and the $\tau$-leap method use a time step $\tau=0.2$, i.e.\ we take eight steps in both methods. The dimension of the problem is therefore 16 (two reaction channels and eight time steps), which is generally thought to be within the realm of possibilities with (R)QMC methods. 

We can clearly see that RQMC applied to both $\tau$-leap and the CLE gives a strong improvement over the same method with standard pseudo random numbers. However, it is also clear that, contrary to the MC method, where both the CLE-based discretisation and $\tau$-leap show equal convergence in terms of the RMSE, the RQMC method shows a difference in performance benefit. The SDE based method has a convergence rate of \highlight{roughly $\mathcal{O}(N^{-1})$} for all $N$. The same behaviour is not observed, however, for the $\tau$-leap method which starts at an $\mathcal{O}(N^{-1})$ rate, but for $N\gtrsim 10^2$ seems to switch to the standard MC rate $\mathcal{O}(N^{-1/2})$. This might come as a surprise, because in the regime of high molecule numbers and reaction propensities the CLE and derived methods are expected to form an excellent approximation to the RTCR and $\tau$-leap method. 

We note that the decrease in convergence rate is not due to sample paths reaching low molecule numbers, which could result in a discrepancy between CLE-based methods and discrete molecule number methods such as the $\tau$-leap method. With the initial conditions given above such sample paths are very unlikely to happen and were not observed in the simulations used to produce \cref{fig:autocatalytic}. This also means that a strategy to prevent negative molecule numbers, e.g.\ \cite{Anderson2008,Cao2005,Tian2004,Chatterjee2005}, was not needed for this example.

\highlight{A clear difference between the $\tau$-leap method and CLE-based method stems from their respective update formulas, \eqref{eq:tau-leap} and \eqref{eq:Langevin}, which are related but not equal. Therefore the results from the two methods can differ subtly. By increasing the reaction rate parameters of the system the Poisson updates used for $\tau$-leaping are better approximated by normal random variables, which is what is used in CLE-based methods. Furthermore, as a result of the difference in updates, the state space of the variable $\boldsymbol{\mathbf{X}}(t)$ is continuous for the CLE-based methods and discontinuous, only taking integer values, for the RTCR-based $\tau$-leap method. We now investigate what differences between the $\tau$-leap method and CLE-based method exactly lead to the two contrasting convergence rate behaviours observed in \cref{fig:autocatalytic}.}

\begin{figure}[h!]
\begin{subfigure}[b]{0.48\textwidth}
\includegraphics[width=\textwidth]{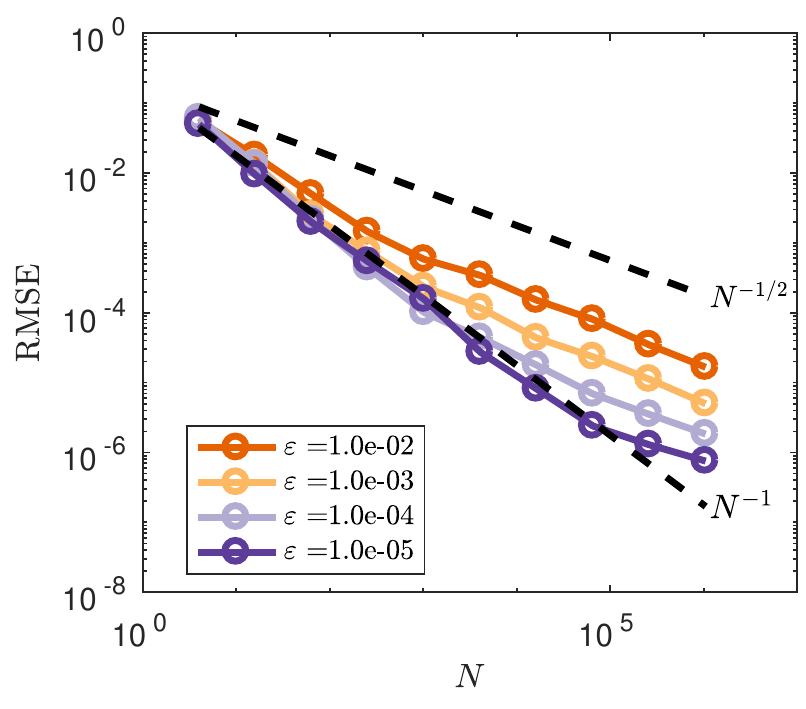}
\end{subfigure}
~
\begin{subfigure}[b]{0.48\textwidth}
\includegraphics[width=\textwidth]{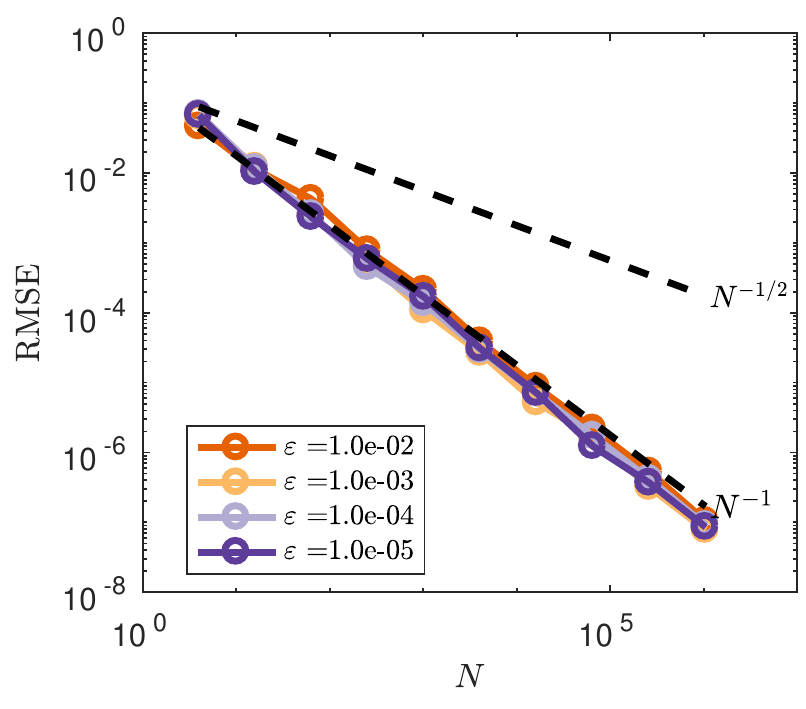}
\end{subfigure}
\caption{Comparison between the normalised RMSE convergence rate between $\tau$-leap (\textit{left}) and an EM discretisation of the CLE (\textit{right}) for the mean number of $S_1$ molecules in \eqref{eq:birth-death} with $c=1$ at $T=1.6$ and varying initial condition $\vect{X}(0)=\varepsilon^{-1}$. Dotted lines show the typical reference convergence rates, $\mathcal{O}(N^{-1/2})$  for MC and $\mathcal{O}(N^{-1})$ for RQMC.}
\label{fig:epsilon-autocatalytic}
\end{figure}

Firstly we test whether the closeness of the $\tau$-leap method and the equivalent discretisation of the CLE changes this observed behaviour of switching between convergence regimes. This is done by running a similar set of simulations with varying initial conditions, and therefore molecule number regimes. We set $\vect{X}(0) = \varepsilon^{-1}$, so that as $\varepsilon\to 0$ we expect better agreement between the $\tau$-leap method and the CLE method. Note that as we vary $\varepsilon$ the sample path variance for both methods has the form $\mathbb{V}\left[\vect{X}(t)\right]=2ct\varepsilon^{-1}$ and therefore grows as $\varepsilon\to 0$. In \cref{fig:epsilon-autocatalytic} we show the resulting comparison between the two methods, with the RMSE rescaled by $\varepsilon^{-1/2}$. This is done to normalise the RMSE by the sample path variance as $\varepsilon$ is changed. Note that this rescaling does not influence the convergence rate behaviour as a function of $N$.

It is clear from \Cref{fig:epsilon-autocatalytic} that for the EM discretisation of the continuous CLE the value of $\varepsilon$ does not influence the convergence rate of the RMSE, i.e.\ it remains $\mathcal{O}(N^{-1})$ under changes in $\varepsilon$. The same cannot be said for the $\tau$-leap method as now $\varepsilon$ influences the transition between two different convergence regimes, fast $\mathcal{O}(N^{-1})$  and slow $\mathcal{O}(N^{-1/2})$ convergence, respectively. We observe that a smaller $\varepsilon$ means that the transition takes place later, i.e.\ for higher $N$. Note that varying $\varepsilon$ in the context of this system means changing the average copy number of $S_1$ encountered, and with that also the average reaction propensities. As a result, $\varepsilon$ toggles how good the Poisson random variables in the $\tau$-leap method can be approximated by normal variables, and therefore how good the CLE is as an approximation to the discrete dynamics. One might therefore think that RQMC performance depends on the `closeness' of a discrete RTCR system is to its continuous CLE approximation. We now show that this is not necessarily the case. 

\begin{figure}[h!]
\begin{subfigure}[b]{0.48\textwidth}
\includegraphics[width=\textwidth]{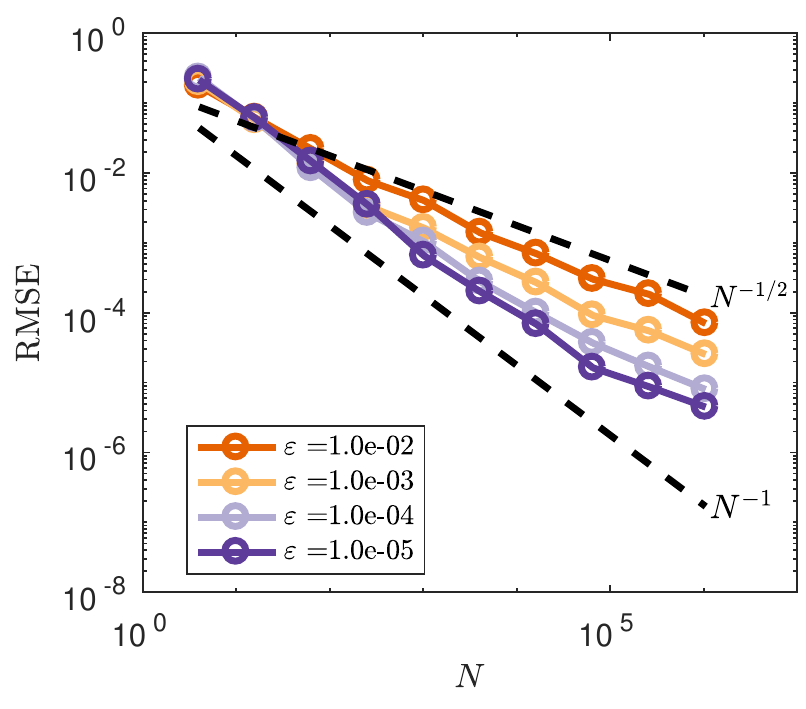}
\end{subfigure}
~
\begin{subfigure}[b]{0.48\textwidth}
\includegraphics[width=\textwidth]{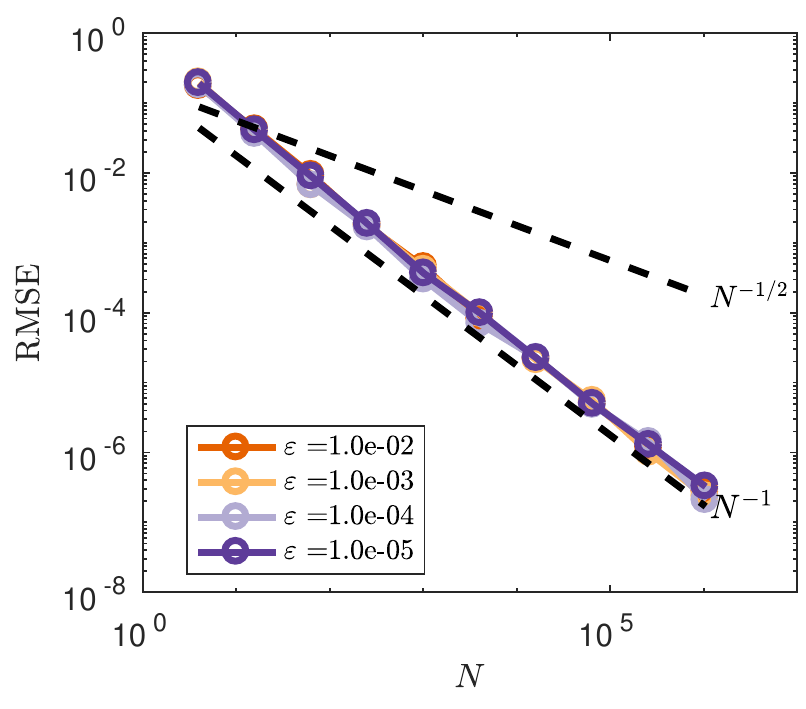}
\end{subfigure}
\caption{Comparison between the normalised RMSE convergence rate between $\tau$-leap (\textit{left}) and an EM discretisation of the CLE (\textit{right}) for the mean number of $S_1$ molecules in \eqref{eq:birth-death} with $c=10\varepsilon$ at $T=1.6$ and varying initial condition $\vect{X}(0)=\varepsilon^{-1}$. Dotted lines show the typical reference convergence rates, $\mathcal{O}(N^{-1/2})$  for MC and $\mathcal{O}(N^{-1})$ for RQMC.}
\label{fig:epsilon-autocatalytic-double}
\end{figure}

We consider an additional rescaling of the reaction rate constant of the form $c=c_0 \varepsilon$ in combination with the previous rescaling of the initial condition. Note that now as $\varepsilon\to 0$ this keeps the reaction propensities on average constant and of the order $\mathcal{O}(c_0\tau)$ during a time step. As a result the value of $\varepsilon$ does not change whether the EM discretisation of the CLE forms a good approximation to the $\tau$-leap method. We perform a test to see what happens to the convergence rate if we change $\varepsilon\to 0$ in this case. The results are shown in \Cref{fig:epsilon-autocatalytic-double} and show similar behaviour compared to the previous example, where $c$ was fixed. It is therefore not a `closeness' of the RTCR to the CLE which governs the convergence rate, as this is determined by the propensities of the reaction channels. Rather it seems to be the copy number of $S_1$ molecules that is crucial for this system.

\subsubsection*{Reversible isomerisation system} The previous example showed that in the case of molecule numbers in the system being not too small RQMC in combination with $\tau$-leaping performed well. In the following example we show that having large molecule numbers for some species in the system, does not guarantee good convergence behaviour of RQMC in combination with $\tau$-leaping. We consider the two species system
\begin{subequations}\label{eq:isomerisation}
\begin{align}
S_1 &\xrightarrow{c} S_2,\\
S_2 &\xrightarrow{\alpha c} S_1,
\end{align}
\end{subequations}
and start with $\vect{X}(0) = (X_1,X_2)^\intercal$ initial molecules. Define $N_0 = X_1+X_2$ and note that this simple system is closed, which means that the sum of the number of $S_1$ and $S_2$ molecules at all times will be equal to $N_0$. This information can be used to decouple the dynamics of $S_1$ and $S_2$. Note that this system, under the CTMC model, converges to an equilibrium state of $(\alpha/(1+\alpha),1/(1+\alpha))^\intercal N_0$. In order to ignore a transient regime in which the system goes to this equilibrium we start the simulations with $N_0=\alpha^{-1}(1+\alpha)\varepsilon^{-1}$ and $\vect{X}(0)$ proportional to this equilibrium state, i.e.\  $\vect{X}(0)= \varepsilon^{-1}(1,\alpha^{-1})^\intercal$.

We note that under this $\vect{X}(0)$ initial condition for both the $\tau$-leap method and the EM discretisation of the CLE we have $\Ex{\vect{X}(t)}=\vect{X}(0)$ and $\mathbb{V}\left[\vect{X}(t)\right]\propto \vect{X}(0)$, like we saw in the previous system. This also means that both computational methods are unbiased for this system.

\begin{figure}[h!]
\includegraphics[width=0.65\textwidth]{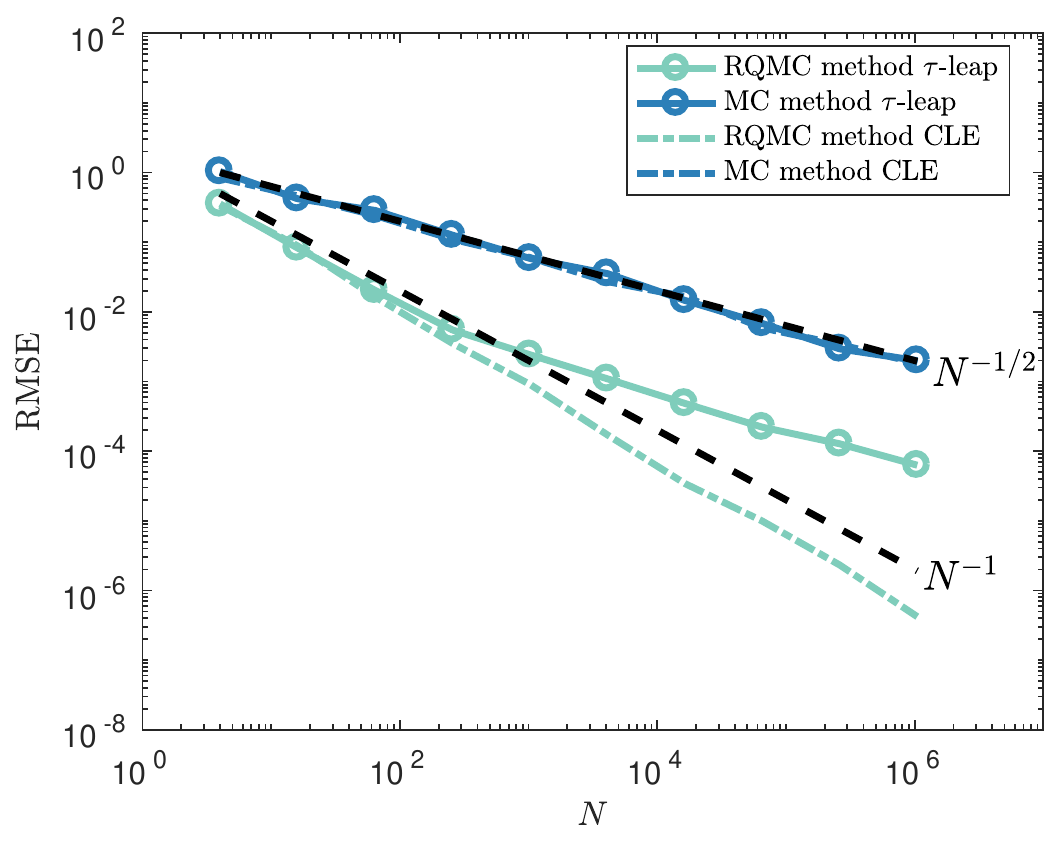}
\caption{RMSE convergence for the mean number of $S_2$ molecules in \eqref{eq:isomerisation} with $c=1$ and $\alpha=10^{-4}$ at $T=1.6$. The time step was $\tau=0.2$ in all simulations. To establish the RMSE \cref{eq:var-QMC} was used with $M=32$ randomisations, both for the RQMC and MC methods. Dotted lines show the typical reference convergence rates, $\mathcal{O}(N^{-1/2})$  for MC and $\mathcal{O}(N^{-1})$ for RQMC.}
\label{fig:isomerisation}
\end{figure}

In \Cref{fig:isomerisation} we show the results for a simulation until $T=1.6$ with time step $\tau=0.2$ and parameters $c=1, \alpha=10^{-4}$ and $\varepsilon=10^{-2}$. This means that $S_2$ has copy numbers of the order $10^6$, which one might reasonably say is large. We note again that there is a gain in performance in terms of RMSE if we compare RQMC and the equivalent MC method.
However, we observe that, despite $S_2$ having large copy numbers, the RMSE for $S_2$ from $\tau$-leaping quickly goes to $\mathcal{O}(N^{-1/2})$ convergence.

\highlight{We can understand this quick slow down of convergence by again noting that $N_0=X_1+X_2$ remains constant. Therefore, the dynamics of $X_2$, and thus also the RMSE, is slaved to the dynamics of $S_1$ molecules (and vice versa), i.e.\ $\text{RMSE}(X_1)=\text{RMSE}(X_2)$. The RMSE of $X_1$ will attain a $\mathcal{O}(N^{-1/2})$ convergence rate relatively quick, because the number of $S_1$ molecules is moderate (on the order of $10^2$), rather than large. The RMSE of $S_2$ molecules mimics this behaviour, because of the coupling via reactions between $S_1$ and $S_2$ molecules, and will therefore also change to $\mathcal{O}(N^{-1/2})$ convergence after the same number of samples $N$.}

This example therefore shows that, by virtue of species being linked through reaction channels, it can be that high copy numbers for part of the reacting species do not guarantee faster convergence rates for RQMC methods than the standard $\mathcal{O}(N^{-1/2})$ rate. \highlight{This is even the case when we use summary statistics that involve just those high copy number species (in our example the number of $S_2$ molecules).}

\subsection*{Discrete toy model}
\begin{figure}[h!]
\begin{subfigure}[b]{0.47\textwidth}
\includegraphics[width=\textwidth]{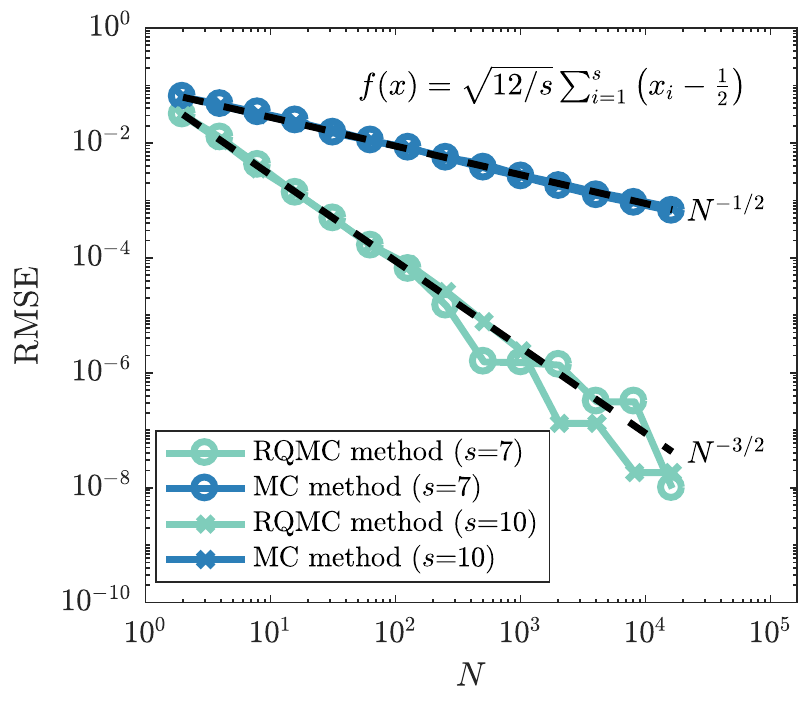}
\end{subfigure}
~
\begin{subfigure}[b]{0.47\textwidth}
\includegraphics[width=\textwidth]{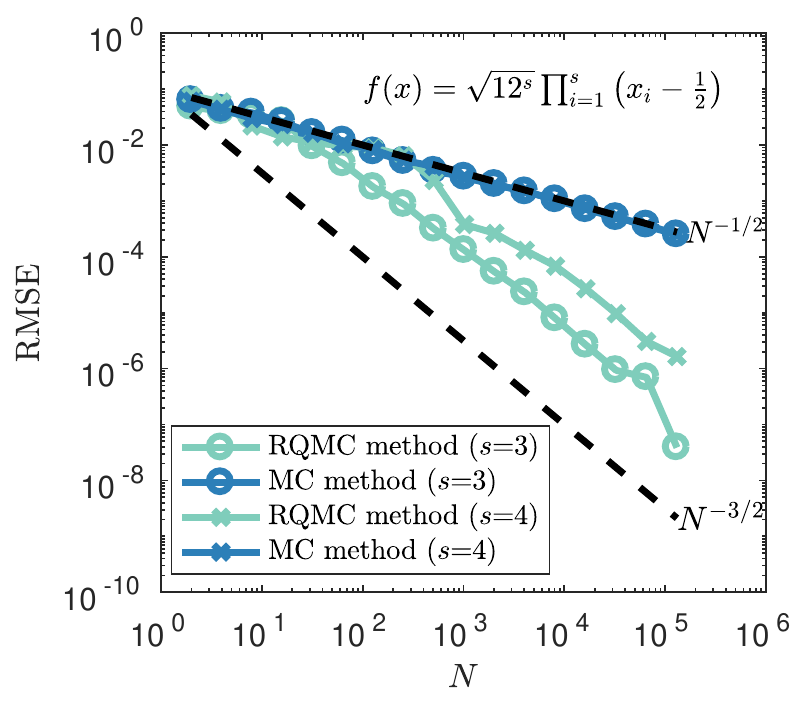}
\end{subfigure}
\caption{Comparison between the RMSE convergence rate of MC and RQMC for \eqref{eq:test-functions}. $M=128$ randomisations were used and dotted lines show the typical reference convergence rates, $\mathcal{O}(N^{-1/2})$  for MC and $\mathcal{O}(N^{-3/2})$ for RQMC.}
\label{fig:testfunctions}
\end{figure}

To explain the observations from the previous examples we consider a problem in traditional quadrature. We consider the integration of the following $s$-dimensional test functions over the domain $[0,1)^s$:
\begin{subequations}\label{eq:test-functions}
\begin{align}
f(x) &= \sqrt{12/s} \sum_{i=1}^s \left(x_i-\frac{1}{2} \right); \label{eq:sum-test}
\\
f(x) &= \sqrt{12^{s}} \prod_{i=1}^s \left(x_i-\frac{1}{2} \right).
\label{eq:prod-test}
\end{align}
\end{subequations}
Both functions integrate to zero over the $s$-dimensional hypercube and have variance
\begin{equation}
\int_{[0,1)^s} f^2(x)\id x - \left( \int_{[0,1)^s} f(x)\id x\right)^2 = 1, 
\end{equation}
regardless of $s$. We note that \eqref{eq:sum-test} is an easy test function for (R)QMC methods as it represents a linear combination of one-dimensional functions (for which (R)QMC methods perform well). The effective dimension in the superposition sense of these additive functions is equal to one \cite{Caflisch1997} and the convergence rate for RQMC\footnote{\highlight{Provable results on the convergence rate for randomised Sobol' sequences are only available if Owen nested uniform scrambling is used \cite{Owen1995}, rather than the randomised matrix scrambling as used in this paper.}} is $\mathcal{O}(N^{-3/2})$ regardless of dimension $s$. The second function \eqref{eq:prod-test} was considered previously in \cite{Owen1998} and is a much harder integrand for RQMC and MC methods. It has the property that RQMC methods for a low number of points have $\mathcal{O}(N^{-1/2})$ RMSE convergence which turns into $\mathcal{O}(N^{-3/2})$ if sufficiently many points are used (the definition of sufficient, which depends on $s$, is found in \cite{Owen1998}). RMSE convergence for these test functions for some dimensions $s$ is depicted in \Cref{fig:testfunctions}. This shows that RQMC does indeed do a very good job at integrating \eqref{eq:sum-test} and for $N$ large enough the same holds for \eqref{eq:prod-test}. For \eqref{eq:sum-test} we see that in terms of RMSE convergence there is no dependency on $s$.

Note that for the chemical test systems previously discussed there was a clear difference in performance for RQMC methods between the continuous CLE and the discrete RTCR. In terms of quadrature, the integrand $f$ in the first case is continuous, whereas in the second case it is discontinuous. Most convergence results for RQMC are based on the assumption that the integrand is continuous and it has been observed before that discontinuities can have an adverse effect on the convergence rate \cite{Morokoff1995,Moskowitz1996,Berblinger1997,He2015}. We now show that a certain type of discontinuity, closely resembling the chemical reaction system case, can replicate the convergence behaviour that we have observed in the previous section.

\begin{figure}[h!]
\begin{subfigure}[b]{0.47\textwidth}
\includegraphics[width=\textwidth]{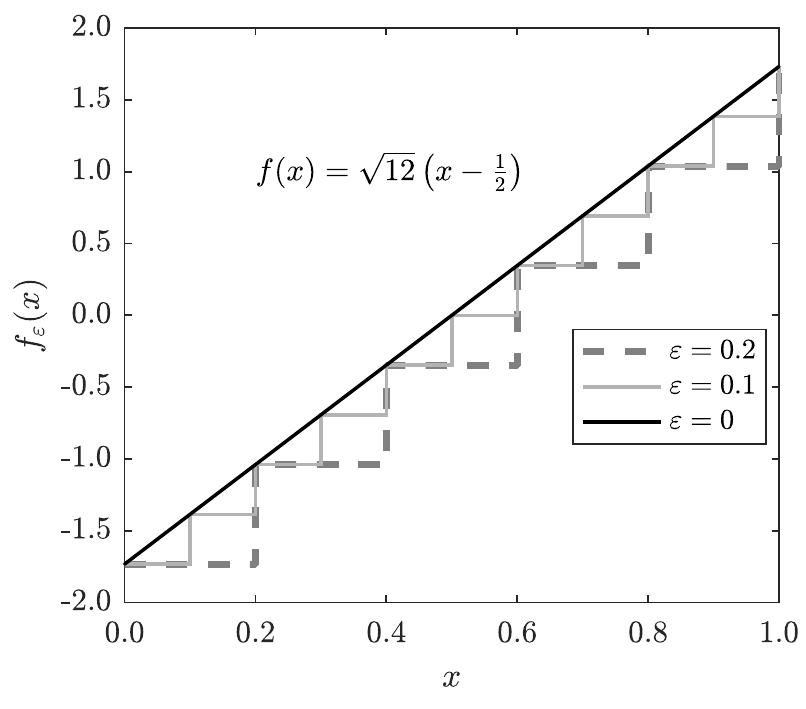}
\end{subfigure}
~
\begin{subfigure}[b]{0.47\textwidth}
\includegraphics[width=\textwidth]{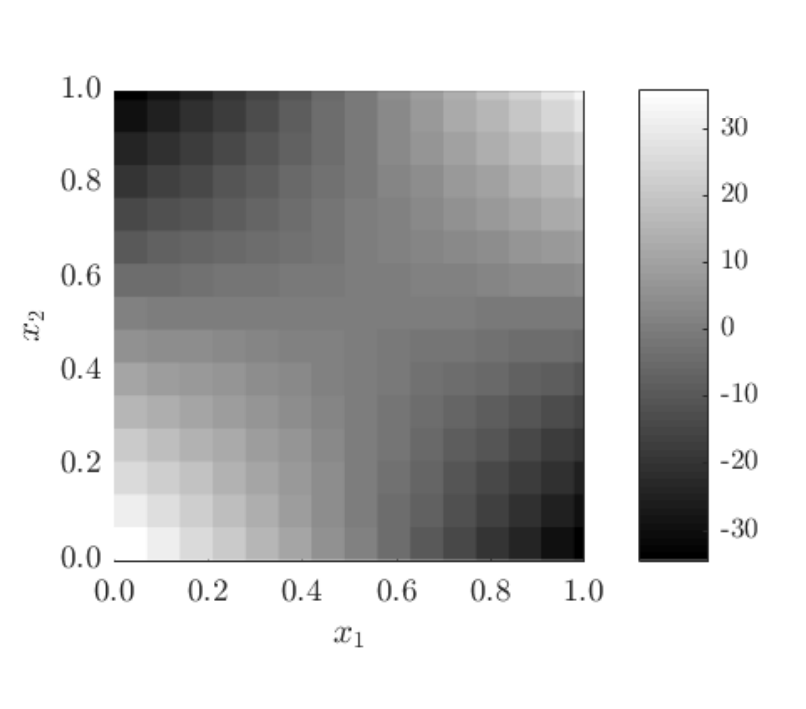}
\end{subfigure}
\caption{Result of the discontinuity transformation \eqref{eq:discontinput}. For the one-dimensional function \eqref{eq:sum-test} we plot the graph of $f_{\varepsilon}(x)$ (\textit{left}). For the two-dimensional function \eqref{eq:prod-test} we plot the filled contour plot for $\varepsilon=0.07$, \highlight{clearly} showing the discontinuity lines of $f_{\varepsilon}(x)$.}
\label{fig:discontfunc}
\end{figure}

We introduce the following transformation of the test functions $f$, which acts upon the input of the function $f$,
\begin{equation}\label{eq:discontinput}
f_{\varepsilon}(x) = f\left( \varepsilon \left\lfloor \frac{x}{\varepsilon}\right\rfloor\right),
\end{equation}
where $\varepsilon$ is a parameter which tunes the level of discontinuity. Note that as $\varepsilon\to 0$ the function becomes smoother. In \Cref{fig:discontfunc} we show the effect of varying $\varepsilon$ on the one-dimensional function \eqref{eq:sum-test} and the filled contour plot for \eqref{eq:prod-test} for $\varepsilon=0.07$. Note that by applying transformation \eqref{eq:discontinput} we create a function which has discontinuities parallel to the axes of the integration domain $[0,1)^s$. \highlight{In \cite{He2015} it was proven that such axes-parallel discontinuities have a relatively mild effect on the convergence of RQMC methods. }

\begin{figure}[h!]
\begin{subfigure}[b]{0.47\textwidth}
\includegraphics[width=\textwidth]{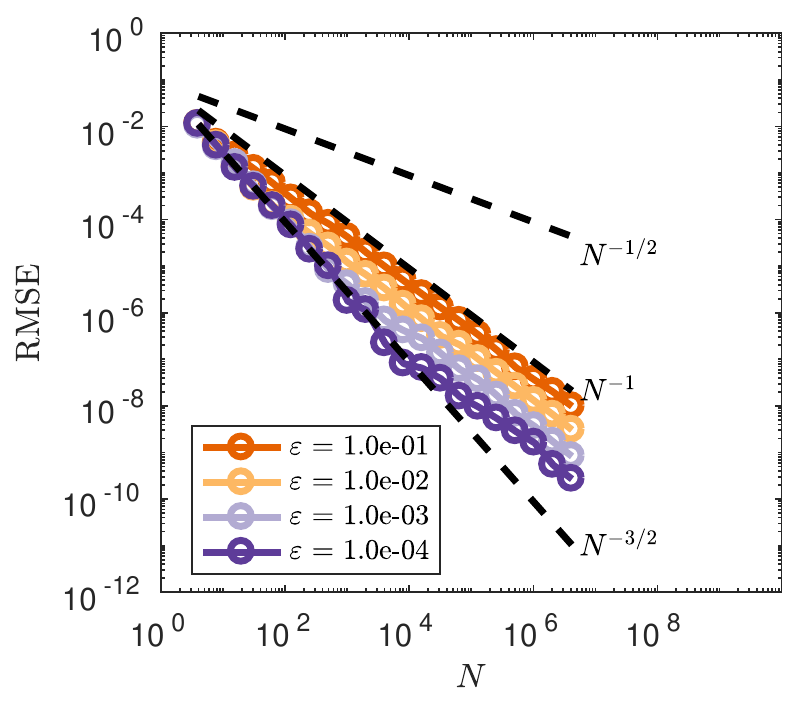}
\end{subfigure}
~
\begin{subfigure}[b]{0.47\textwidth}
\includegraphics[width=\textwidth]{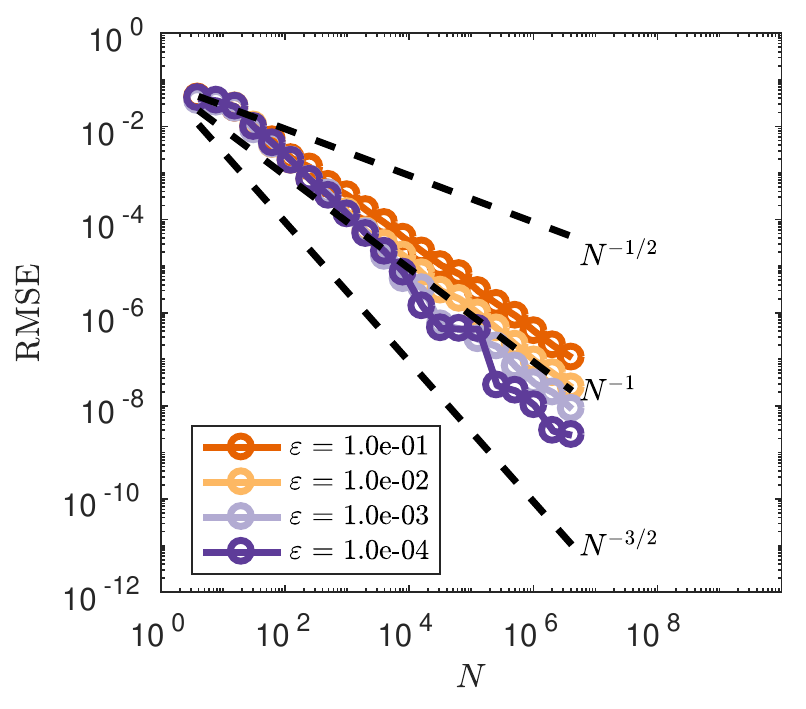}
\end{subfigure}
\caption{Effect of discontinuity transformation \eqref{eq:discontinput} on the RMSE convergence for \eqref{eq:sum-test} ($s=10$) and \eqref{eq:prod-test} ($s=3$). $M=128$ randomisations were used and dotted lines show typical reference convergence rates, $\mathcal{O}(N^{-1})$ and $\mathcal{O}(N^{-3/2})$ for RQMC.}
\label{fig:testfunctions-discontinput}
\end{figure}

In \Cref{fig:testfunctions-discontinput} we see the effect that the introduction of discontinuity by \eqref{eq:discontinput} has on the RMSE convergence. Where the continuous functions showed $\mathcal{O}(N^{-3/2})$ convergence (recall \Cref{fig:testfunctions}), the discontinuous counterparts have, for large enough $N$, a slower $\mathcal{O}(N^{-1})$ convergence rate. The results in \Cref{fig:testfunctions-discontinput} hold for a wide range of dimensions $s$. As expected, results for \eqref{eq:sum-test} are not affected by $s$ due the fact that the function after transformation still is one-dimensional in superposition sense. On the other hand for \eqref{eq:prod-test} the effect of transformation \eqref{eq:discontinput} only shows once enough points have been used to leave the $\mathcal{O}(N^{-1/2})$ initial convergence, and after that convergence rates seem to drop from $\mathcal{O}(N^{-3/2})$ to $\mathcal{O}(N^{-1})$ as well.

\begin{figure}[h!]
\begin{subfigure}[b]{0.4\textwidth}
\includegraphics[width=\textwidth]{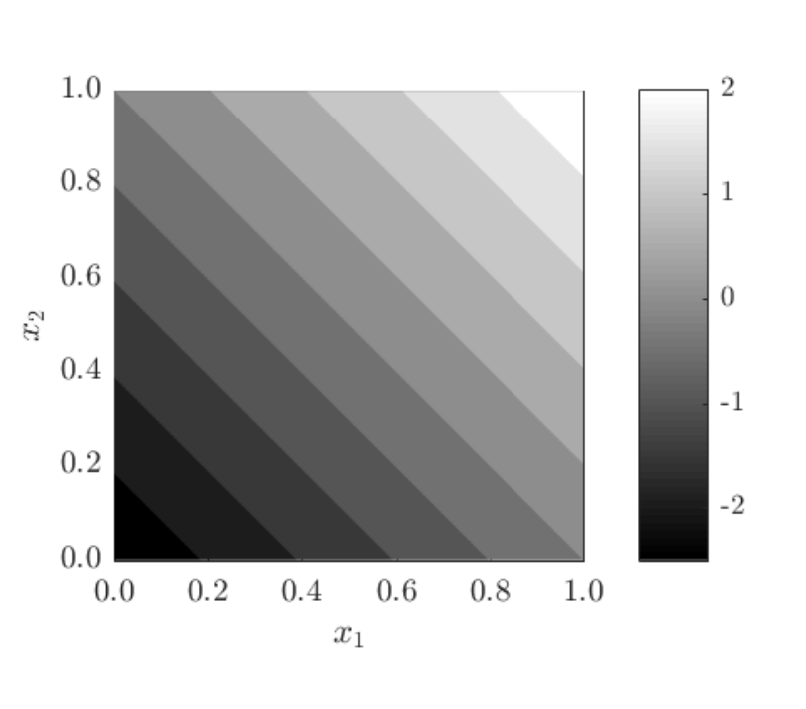}
\end{subfigure}
~
\begin{subfigure}[b]{0.4\textwidth}
\includegraphics[width=\textwidth]{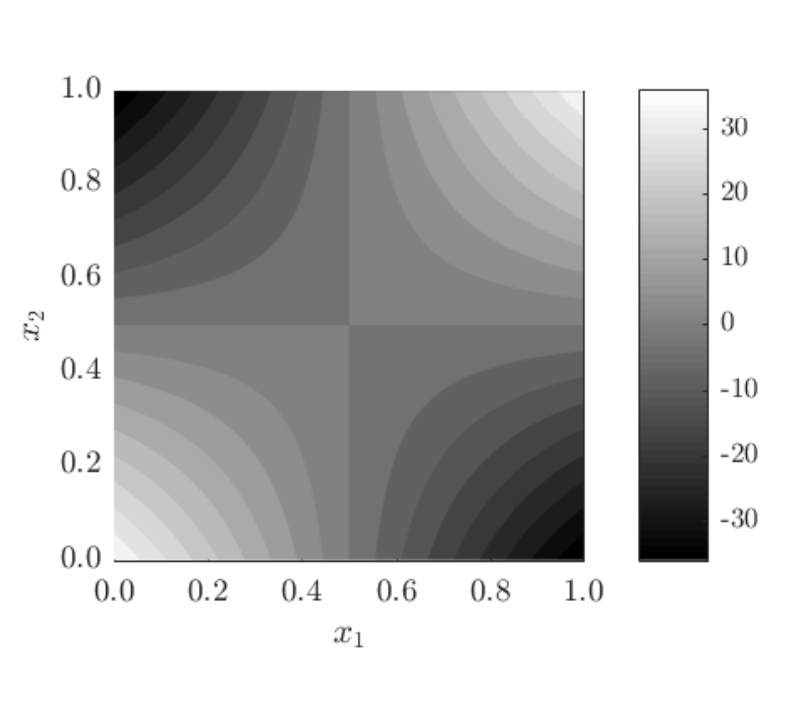}
\end{subfigure}
\caption{Result of the discontinuity transformation \eqref{eq:discontoutput}. Plots show the filled contour plot for \eqref{eq:sum-test} with $\varepsilon=0.5$ (\textit{left}) and the filled contour plot for \eqref{eq:prod-test} with $\varepsilon=4$ (\textit{right}). The discontinuity lines of $f_{\varepsilon}(x)$ do not align with the axes of $[0,1)^2$.}
\label{fig:2ddiscontfunc}
\end{figure}

Next we introduce a different transformation that converts continuous functions into discontinuous ones,
\begin{equation}\label{eq:discontoutput}
f_{\varepsilon}(x) = \varepsilon \left\lfloor \frac{f(x)}{\varepsilon}\right\rfloor.
\end{equation}
Note that, in contrast to \eqref{eq:discontinput}, this transformation acts upon the function output values. As a result, discontinuities introduced by \eqref{eq:discontoutput} do not necessarily align with the axes of $[0,1)^s$, as can be seen in \Cref{fig:2ddiscontfunc} in two dimensions. 

\begin{figure}[h!]
\begin{subfigure}[b]{0.47\textwidth}
\includegraphics[width=\textwidth]{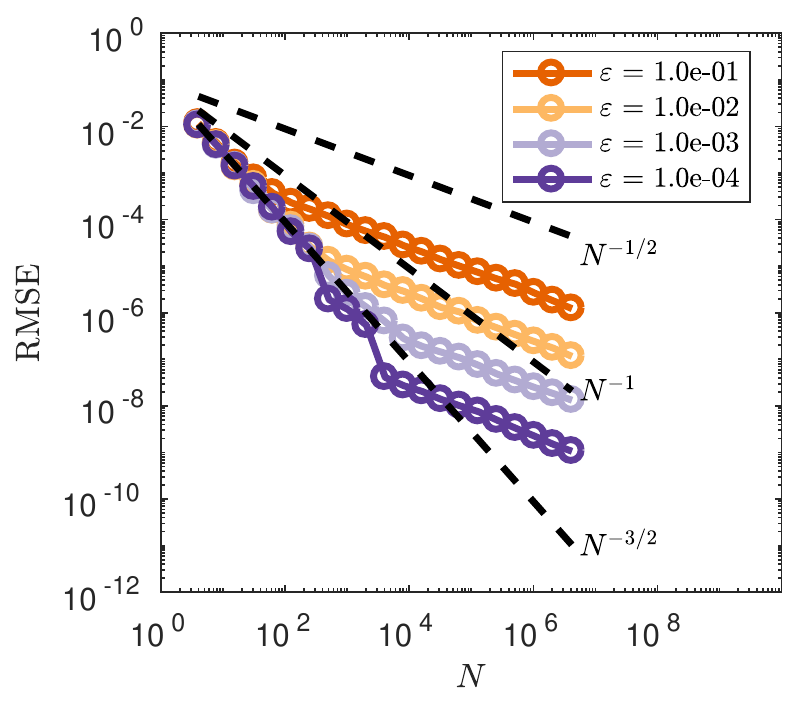}
\end{subfigure}
~
\begin{subfigure}[b]{0.47\textwidth}
\includegraphics[width=\textwidth]{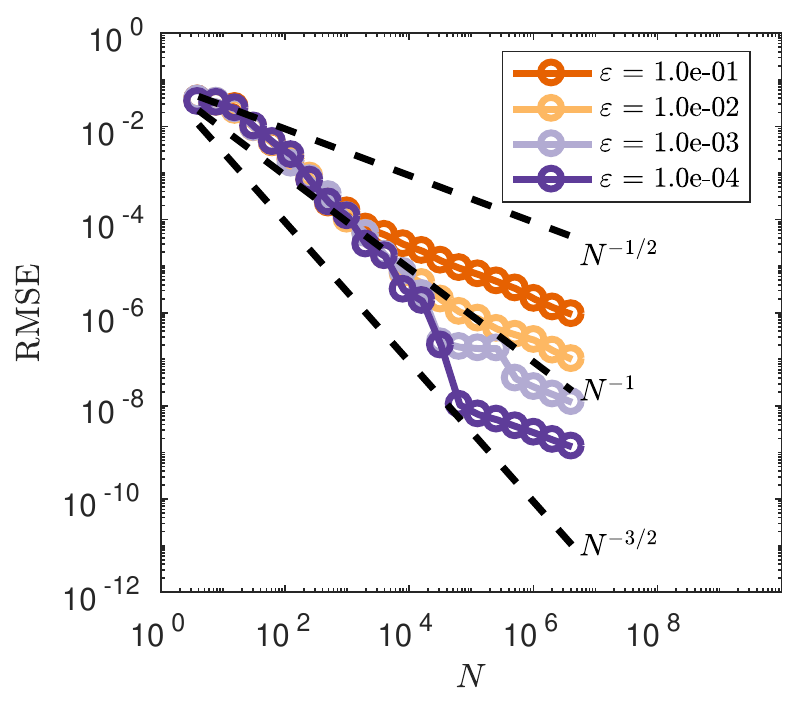}
\end{subfigure}
\caption{Effect of discontinuity transformation \eqref{eq:discontoutput} on the RMSE convergence for \eqref{eq:sum-test} ($s=10$) and \eqref{eq:prod-test} ($s=3$). $M=128$ randomisations were used and dotted lines show typical reference convergence rates, $\mathcal{O}(N^{-1})$ and $\mathcal{O}(N^{-3/2})$ for RQMC.}
\label{fig:testfunctions-discontoutput}
\end{figure}

Results for the RMSE convergence for varying $\varepsilon$ is shown in \Cref{fig:testfunctions-discontoutput}. We observe again that for small values of $N$ the convergence rate is $\mathcal{O}(N^{-3/2})$, similar to the continuous case. However, we see that with transformation \eqref{eq:discontoutput}, for $N$ large enough, the convergence rate becomes $\mathcal{O}(N^{-1/2})$, rather than $\mathcal{O}(N^{-1})$ which was observed for transformation \eqref{eq:discontinput}. This comes back to the fact that the discontinuities introduced by \eqref{eq:discontoutput} do not align with the axes of the integration domain $[0,1)^s$. \highlight{One can understand this from the way many RQMC point sets are constructed (in particular digital nets, of which Sobol' point sets are a special case). For such sets the points are equidistributed with respect to axes-aligned hyperrectangles. If the discontinuities of the integrand do not align with the domain axes, such as for transformation \eqref{eq:discontoutput}, then the RQMC points will not be able to sample of the integrand's different contributions uniformly. Discontinuities that do not align with the domain axes were also shown to be of a more detrimental type of discontinuity if one wants to use RQMC methods in \cite{He2015}. }

The limiting convergence rate is given by the MC rate $\mathcal{O}(N^{-1/2})$. This agrees with the fact that RQMC methods will, in the worst case scenario, behave very much like a standard MC method and have a convergence rate which is not more than a constant times the MC rate \cite{Owen1998}.

To further explain the convergence behaviour we consider the decomposition of the discontinuous function into a continuous part, $F(x)$, and discontinuous part, $G(x)$, of the form
\begin{equation}
f_{\varepsilon}(x) = \underbrace{f(x)}_{\text{continuous } F(x)} + \underbrace{\left(f_{\varepsilon}(x)-f(x) \right)}_{\text{discontinuous } G(x)}.
\end{equation}
Note that $|G(x)|\leq \varepsilon$ and as a result the variance of $G(x)$ is generally $\mathcal{O}(\varepsilon^2)$. We can then decompose the MSE of the estimator of the integral of $f_{\varepsilon}(x)$ by an unbiased RQMC rule as the sum of the MSE of the integration of $F(x)$ and $G(x)$. We note that the MSE for the continuous part, $F(x)$, behaves like $\mathcal{O}(N^{-3})$, as observed in \Cref{fig:testfunctions}. In the case of transformation \eqref{eq:discontoutput} the RQMC method achieves MC like error rates for the discontinuous part, $G(x)$. We therefore have the following decomposition of the MSE
\begin{equation}
\text{MSE}\left( \varepsilon \left\lfloor \frac{f\left(x\right)}{\varepsilon}\right\rfloor \right) = C_1 N^{-3} + C_2 \varepsilon^2 N^{-1}.
\end{equation}
This yields a switch from fast $\mathcal{O}(N^{-3})$ convergence to slow $\mathcal{O}(N^{-1})$ when $N=\mathcal{O}(\varepsilon^{-1})$, i.e.\ at this point the error made for the discontinuous component of the function dominates the MSE. The same holds true for the RMSE and this scaling of the switch point as a function of $\varepsilon$ is also observed in \Cref{fig:testfunctions-discontoutput}.

In the case of transformation \eqref{eq:discontinput} the RQMC method does not perform like a MC method and instead achieves $\mathcal{O}(N^{-2})$ convergence for the MSE. Note that the scaling of the variance now does not come into play, because the convergence is not MC-like. Instead we observe a rescaling of the switching point $N=\mathcal{O}(\varepsilon^{-1})$ as well in \Cref{fig:testfunctions-discontinput}. This leads to the following decomposition of the MSE
\begin{equation}
\text{MSE}\left( f\left( \varepsilon \left\lfloor \frac{x}{\varepsilon}\right\rfloor \right)\right) = C_1 N^{-3} + C_2 \varepsilon N^{-2}.
\end{equation}

This shows that, even in the case of a discontinuous integrand, RQMC methods can achieve lower MSE if the function can be decomposed in a continuous part and a discontinuous part that is relatively smaller in magnitude. RQMC performs superiorly on the continuous component of the integrand, giving fast error decay for a moderate number of points $N$. In the worst-case scenario a MC convergence rate is achieved by RQMC on the discontinuous part, which will dominate the convergence order for large $N$.

This observation can be linked to observations made in \cite{Caflisch1998}. Caflisch notes that low-discrepancy point sets differ subtly from pseudo random point sets in the sense that for a pseudo random point set every point is an independent estimate of the integral. This is not true for a low-discrepancy point set, which has a deterministic structure. For these point sets the initial points sample the integration domain on a coarse scale, whereas the later points are used for progressively finer scales. Therefore initially RQMC will perform well on a function like $f_{\varepsilon}$, because on a coarse scale it is dominated by its continuous part, $F(x)$. If more points are used the fine, discontinuous, structure due to $G(x)$ starts to dominate and this is where the convergence stalls.

Note that for a general chemical reaction network it is not clear \textit{a priori} how the summary statistic of interest can be decomposed into a continuous part and discontinuous part, or what the value of $\varepsilon$ is. Or, in other words, it is not clear how important coarse scale continuous contributions are in relation to finer scale discrete ones. Therefore the performance benefit from using RQMC methods over MC methods can be hard to estimate \textit{a priori}. We do, however, note that the implementation of low-discrepancy point sets is often relatively simple and does not need to increase the runtime of the simulation procedures (\Cref{ap:qmc-timing}). As a result, RQMC methods have a potential to provide computational savings over MC methods in the simulation approaches of chemical reaction networks by attaining lower statistical errors for similar computational time.

\subsection*{Schl\"ogl system}
As a final example we look at the bistable Schl\"ogl system, as encountered in \cite{Cao2004}, which incorporates non-linear interactions
\begin{subequations}\label{eq:schlogl}
\begin{align}
2 S_1 + S_2 &\xrightarrow{c_1} 3 S_1,\\
3 S_1 &\xrightarrow{c_2} 2 S_1 + S_2,\\
S_3 &\xrightarrow{c_3} S_1,\\
S_1 &\xrightarrow{c_4} S_3,
\end{align}
\end{subequations}
where we assume that the copy numbers for $S_2$ and $S_3$ are constant and large. The initial condition for $S_1$ is 250 molecules. Non-dimensional parameters are given by $c_1=3\cdot10^{-7}$, $c_2 = 10^{-4}$, $c_3=10^{-3}$, $c_4 = 3.5$ and the copy numbers for $S_2$ and $S_3$ are taken as $10^5$ and $2\cdot 10^5$, respectively. The system is bistable for these parameters, with stable states around 100 copy numbers and 550 copy numbers for $S_1$.

We simulate the system up until final time $T=4$ with time step $\tau=0.4$. We take the approach in \cite{Anderson2012} to deal with sample paths zero or fewer molecule numbers at a given time. We look at the mean number of $S_1$ molecules, though more meaningful summary statistics can be constructed for bistable systems.

\begin{figure}[h!]
\includegraphics[width=0.65\textwidth]{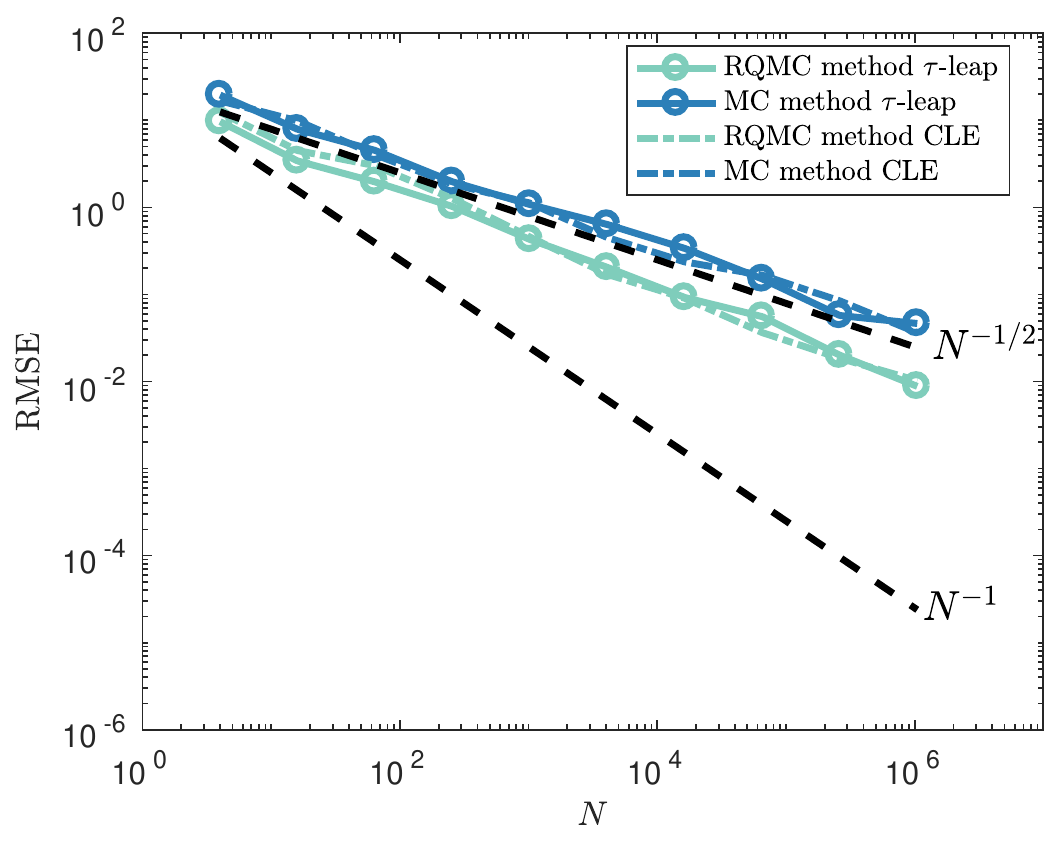}
\caption{RMSE convergence for the mean number of $S_1$ molecules in \eqref{eq:schlogl}. The time step was $\tau=0.4$ in all simulations. To establish the RMSE \cref{eq:var-QMC} was used with $M=32$ randomisations, both for the RQMC and MC methods. Dotted lines show the typical reference convergence rates, $\mathcal{O}(N^{-1/2})$  for MC and $\mathcal{O}(N^{-1})$ for RQMC. \highlight{Estimated convergence rate for RQMC methods is $\mathcal{O}(N^{-\nu})$ with $\nu\approx 0.55$.}}
\label{fig:schlogl}
\end{figure}

In \Cref{fig:schlogl} we show results comparing the $\tau$-leap method and EM discretisation of the CLE using both pseudo random points and low-discrepancy points. We see that, although the RQMC method does not attain a much higher convergence rate than the standard MC rate of $\mathcal{O}(N^{-1/2})$, the RQMC method is superior to the standard MC method. Numerical experiments suggest that a similar situation as in \Cref{fig:schlogl} holds for at least the first few moments of $S_1$ copy numbers.

We also observe that, even though the CLE is continuous, the convergence rate for the EM discretisation is equal to that of $\tau$-leap. This indicates that for this specific problem it might not be the discrete nature of the $S_1$ dynamics that causes the observed $\mathcal{O}(N^{-1/2})$ convergence rate. The behaviour is likely due to the fact that the system has four reaction channels and 10 time steps, leading to a dimensionality of 40 for this specific problem. Such a number of dimensions can be challenging for na\"ive QMC methods as applied here. One might benefit from applying a change of variables which transforms the effective dimension of the problem, and therefore improves the RQMC convergence. Techniques such as the Brownian bridge and principal components construction are available for SDEs and can help in making RQMC methods effective even for high dimensional problems \cite[Chapter 5]{Glasserman2003}. For dynamics following the RTCR \eqref{eq:KurtzRT} such transformations are, however, not known and we leave this direction for future research.

\section{Discussion and outlook}\label{sec:discussion}
It is known that the use of low-discrepancy numbers instead of pseudo random numbers can greatly improve the convergence speed for problems involving traditional quadrature and SDEs. In this paper we explored the application of RQMC methods in the framework of simulation of stochastic biological systems. In particular, we looked at the combination of low-discrepancy numbers with the $\tau$-leap method. For simplicity, the fixed step $\tau$-leap method was considered so as to allow for a simple implementation of low-discrepancy points without negative effects on the runtime. We note that the question of whether this is a good procedure has been addressed in the literature before \cite{Cao2004,Cao2005,Cao2006,Anderson2008}. This paper, however, does not focus on the question of whether $\tau$-leaping forms a good approximation to the CTMC dynamics, which is the motivation therein for the discussion about time step selection. Rather, we focus on the question of how quick statistical errors in desired summary statistics decay as a function of the number of sample paths simulated. We answer this question in the simplest possible case, namely using fixed time step $\tau$-leap, though we expect our conclusions below to be general enough to hold for a large class of simulation procedures for stochastic biological systems.

Theory suggests that the convergence rate for an RQMC method is not worse than for the equivalent MC method (up to a constant \cite{Owen1998}). Reality seems to show that in case of chemical reaction networks RQMC is superior to MC, as evidenced by numerical experiments in \Cref{sec:results}. \highlight{As a result, if one chooses the fixed time-step $\tau$-leap approach to simulate a chemical reaction network, the use of RQMC methods gives a better convergence behaviour as compared to the traditional MC implementation at no extra cost.} 

However, the benefits from using low-discrepancy numbers are smaller than anticipated based on results seen in the simulation of SDEs. In particular, if one chooses to model chemical reaction systems by SDEs in the form of the CLE, one sees a greater advantage in the use of low-discrepancy numbers. This effect can be caused by at least two factors. 

Firstly, the inherently discrete nature of stochastic simulations of chemical reaction networks hinders RQMC convergence. It has been reported in the literature that discontinuous integrands experience less benefit from RQMC methods over standard MC methods \cite{Morokoff1995,Moskowitz1996,Berblinger1997,He2015}. In \Cref{sec:results} we showed through the use of a simplified test system that the behaviour observed in simulating chemical reactions can be replicated by introducing certain types of discontinuity in classical quadrature. The simple test systems in \Cref{sec:results} allow for a detailed understanding of the RMSE convergence rate observed when applying RQMC. It is, however, not always possible to choose the biological model or its parameters such that the effect of discontinuities will be small. It would therefore be advantageous to have techniques that leave the desired summary statistic intact, but diminish the effect of discontinuities on the RMSE convergence. Smoothing techniques have previously been considered to mitigate the effects of discontinuity in other contexts \cite{Caflisch1998,Moskowitz1996} and we aim to address similar techniques in the context of chemical reaction networks in future work.

Secondly, it is known that the performance of (R)QMC methods can strongly depend on the dimension of the problem. As illustrated in the last example in \Cref{sec:results}, a \highlight{higher} dimension can lead to a much smaller performance benefit, regardless of the smoothness of the underlying problem. Methods to reduce the effective dimension of the problem by a change of variables have proven to be effective in other fields and it is an open question as to whether such transformations can be found for the simulation of biological systems. Another method which has proven to be fruitful in the simulation of DTMCs of potentially large dimension is array-RQMC \cite{LEcuyer2008}. This method is the cornerstone of the only other known QMC work in the area of stochastic biological systems \cite{Hellander2008}. Observations in this paper about the effect of the discrete nature of chemical reaction systems support and explain the observation of a smaller than expected performance gain in \cite{Hellander2008}. In future work, we will explore the effect of discontinuities on the array-RQMC method and its combination with $\tau$-leaping, with both fixed and adaptive time stepping.

We also point out that the original article introducing QMC methods in 1951 by Richtmyer \cite{Richtmyer1951} considered a discrete linear birth process. He observed a smaller performance gain than expected and this might have impeded the further exploration of QMC methods in stochastic simulation for a few decades. Richtmyer's results can now be understood to be caused by the unfortunate choice of his chosen model problem, which is discontinuous in nature.

A further topic of future research is the effective dimension in the simulation of the RTCR. The concept of effective dimension and techniques to reduce said dimension are widely studied in the context of financial applications and in the future we aim to explore its implications for the models of interest in biology.

\subsection*{Acknowledgements}
The authors are grateful to M.\ B.\ Giles for insightful discussions and suggestions made throughout this work. The authors would also like to thank the anonymous reviewers, whose detailed comments improved the clarity of this paper. Casper H.\ L.\ Beentjes acknowledges the Clarendon Fund and New College, Oxford, for funding. Ruth E.\ Baker is a Royal Society Wolfson Research Merit Award holder and a Leverhulme Research Fellow, and also acknowledges the BBSRC for funding via grant no.\ BB/R000816/1.

\newpage
\appendix
\section{Computational effort to generate quasi-random numbers}\label{ap:qmc-timing}
\begin{figure}[h!]
\begin{subfigure}[b]{0.46\textwidth}
\includegraphics[width=\textwidth]{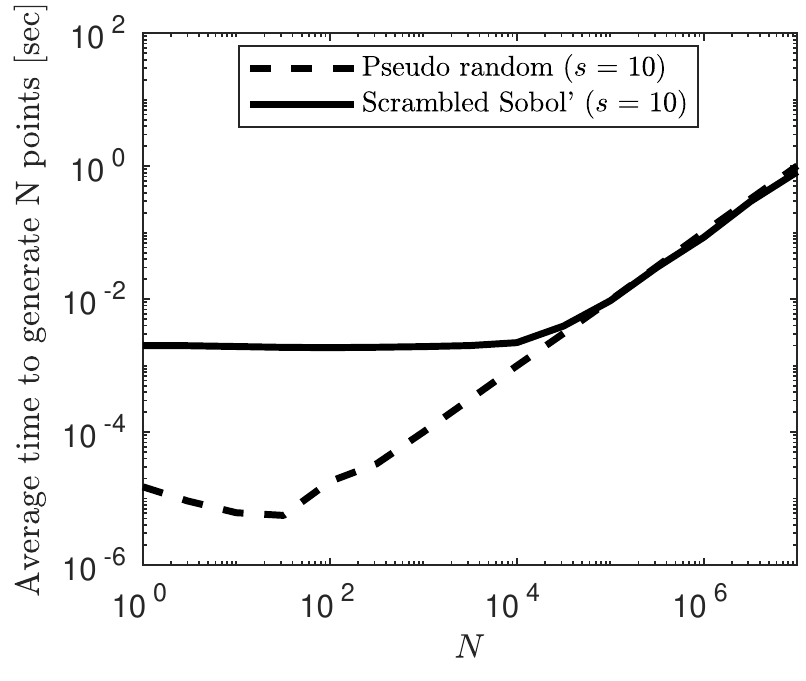}
\end{subfigure}
~
\begin{subfigure}[b]{0.46\textwidth}
\includegraphics[width=\textwidth]{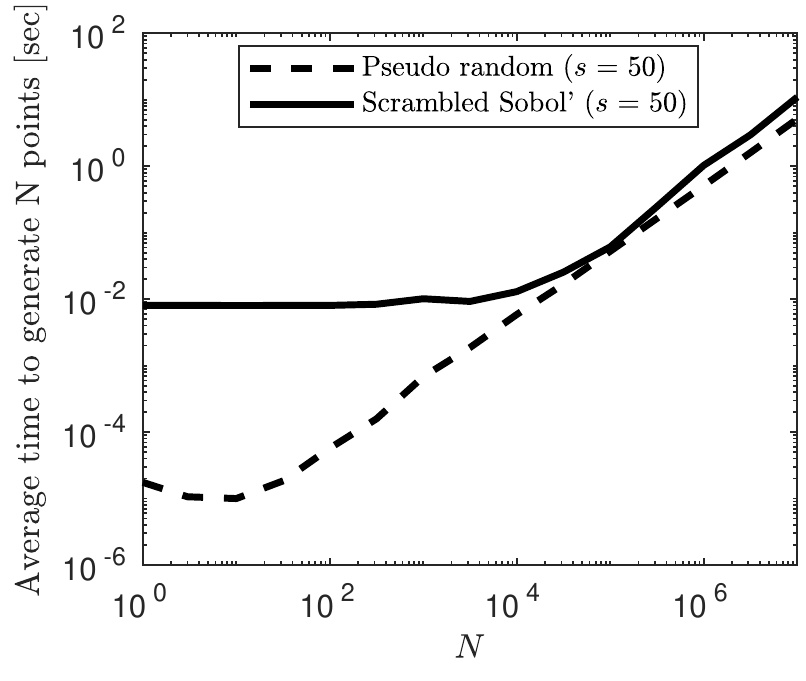}
\end{subfigure}
\caption{Comparison between the time taken to generate $N$ pseudo-random points and an equal number of scrambled Sobol' points. On the left and right results for $s=10$ and $s=50$ dimensional points, respectively.}
\label{fig:timing}
\end{figure}

One should question whether the time taken to generate scrambled low-discrepancy sequences for an RQMC method is much longer than the time needed for pseudo-random numbers to be generated as this could void any observed performance gains. We therefore perform a small test to time the generation of the various random numbers. We time how long it takes to generate a point set of length $N$ in $s$ dimensions (averaged over 50 trials). Timing experiments were performed using MATLAB R2017b on an Ubuntu desktop PC with a 3.40 GHz Intel Core i7-2600K CPU and 16 GB of random access memory. We test the standard pseudo-random number generator (which uses the Mersenne Twister algorithm) versus Sobol' points with \highlight{linear matrix scrambling and a random digital shift}. The results are depicted in \cref{fig:timing} and show that only for relatively small point sets the generation of pseudo-random numbers is distinctly faster than the Sobol' points (on the order of milliseconds). For point sets of lengths not uncommon in simulations ($10^5$ or more points) the difference is negligible. Therefore the completion time for an algorithm which has replaced pseudo-random numbers with low-discrepancy numbers will not differ noticeably. These findings agree with practical timing results for simulations of various financial applications in \cite{Lemieux2009}.

\section{RQMC as a MC variance reduction technique}\label{ap:RQMCvarred}
An alternative look on RQMC is as a variance reduction technique within the standard MC framework as noted in \cite{LEcuyer2016}. After randomisation of the low-discrepancy point set the estimator $I_N^{(m)}$ becomes an unbiased estimator of the integral $I$ in \cref{eq:integral}. The variance of the estimator can, by linearity, be written as
\begin{equation}\label{eq:RQMCvarred}
\mathbb{V}\left[I_N^{(m)}\right] = \frac{\sigma^2}{N} + \frac{2}{N^2}\sum_{1\leq i\leq j\leq N} \text{Cov}\left[f\left(\hat{\vect{v}}^{(i,m)}\right),f\left(\hat{\vect{v}}^{(j,m)}\right) \right].
\end{equation}
In standard MC methods the points $\left\{ \hat{\vect{v}}^{(i,m)} \right\}$ used are independent and therefore the covariances are zero. For an RQMC method, however, this is not the case because of the deterministic construction of the points used. Note that this remains true despite the randomisation, because the point set as a whole still is a low-discrepancy set. In order to reduce the variance, one wants the contribution of the sum of covariances to be as negative as possible. This is attempted by RQMC methods through the construction of the points used in the quadrature and it places RQMC methods on equal footing with, for example, the standard variance reduction techniques of antithetic sampling and common random numbers \cite[Chapter 4]{Lemieux2009}.



\setlength\bibitemsep{4pt}
\printbibliography
\end{document}